\documentclass[11pt]{article}
\usepackage{epsfig,amsmath,latexsym,amssymb}
\usepackage{amscd}
\usepackage{multirow}
\usepackage{graphicx}
\usepackage{lscape}
\def\R{{\mathbb R}}  %%
\def\N{{\mathbb N}}  %%
\def\p{{\mathbb P}}  %% Schoene Darstellungen der Zahlenmengen
\def\E{{\mathbb E}}  %%
\def\F{{\mathbb F}}  %%

\newcommand{\Remm}[1]{}
\newtheorem{theo}{Theorem}[section]
\newtheorem{lemma}[theo]{Lemma}

\newtheorem{model ass}[theo]{Model Assumptions}

\newtheorem{conclusions}[theo]{Conclusions}

\def\EndProof{{\begin{flushright}\vspace{-2mm}$\Box$\end{flushright}}}
\numberwithin{equation}{section}

\def\Beweis{\footnotesize}

\begin{document}
\author{Josef Teichmann\footnote{ETH Zurich, Department of Mathematics, 8092 Zurich, Switzerland} \qquad Mario V.~W\"uthrich$^{\ast}$}

\date{\today}
\title{Consistent Long-Term Yield Curve Prediction}
\maketitle

\begin{abstract}
%\begin{center}
%\noindent
We present an arbitrage-free non-parametric yield curve prediction model which takes the full (discretized) yield curve as state variable. We believe that absence of arbitrage is an important model feature in case of highly correlated data, as it is the case for interest rates. Furthermore, the model structure allows to separate clearly the tasks of estimating the volatility structure and of calibrating market prices of risk. The empirical part includes tests on modeling assumptions, back testing and a comparison with the Vasi\v{c}ek short rate model.
%\end{center}
\end{abstract}

\section{Zero coupon bond prices and yield curves}
\label{section introduction}
Insurance cash flows are valued using the risk-free yield curve. 
First, today's yield curve needs to be estimated 
from government bonds, swap rates and corporate bonds and, second, future 
yield curves then need to be predicted.  This prediction is a complex task because, in general, it involves
the forecast of infinite dimensional random vectors and/or random functions. In the
present paper we tackle the problem of yield curve prediction using a non-parametric approach, which is based on ideas
presented in Ortega et al.~\cite{Ortega}. In contrast to \cite{Ortega} we are heading for long term predictions as needed in insurance industry. Assume $t\ge 0$ denotes time in years. Choose $T \geq t$ and denote, at time $t$, the price of the (default-free) zero coupon bond (ZCB) that pays one unit of currency at maturity date $T$ by $P(t,T)$. The yield curve at time $t$  for maturity dates $T \geq t$ is then given by the continuously-compounded spot
rate defined by \begin{equation*}
Y(t,T) =- \frac{1}{T-t}~\log P(t,T).
\end{equation*}

{\bf Aim and scope.} 

Model stochastically the yield curves $T \mapsto Y(t,T)$ for future dates
$t\in (0,T)$ such that:

(i) the model is free of arbitrage;

(ii) explains past yield curve observations;

(iii) allows to predict the future yield curve development.

In contrast to standard literature on prediction of yield curves we insist that models should be free of arbitrage. This requirement is crucial when it comes to the prediction of highly correlated prices as it is the case for interest rates. Otherwise it is possible to ``artificially'' shift P\&L distributions.  More precisely, if a prediction model admits arbitrage then implementing this arbitrage portfolio yields an always positive P\&L. In practice adding such an arbitrage portfolio can then be used to shift P\&L distributions of general portfolios, which is an undesired effect from the point of view of 
valuation and risk management, see Figure \ref{arbitrage}
and Section \ref{sec.arbitrage}.

~

{\bf Organization of the paper.} 
The remainder of the paper is organized as follows: in Section 2 we propose our discrete time model for (discretized) yield curve evolution. In Section 3 we describe the ubiquitous no arbitrage conditions for our modeling setup. In Section 4 we describe the actual calibration procedure and in Section 5 we present a concrete calibration to real market data.

\section{Model proposal on a discrete time grid} 

Choose a fixed grid size $\Delta = 1/n$ for $n\in \N$. We consider
the discrete time points $t \in \Delta \N_0=\{0,\Delta, 2\Delta, 3\Delta, \ldots\}$ and the maturity dates $T\in t+\Delta \N$. For example, the choice
$n=1$ corresponds to a yearly grid, $n=4$ to a quarterly grid,
$n=12$ to a monthly grid, $n=52$ to a weekly grid
 and $n=250$ to a business days grid.

~

The filtered probability space is denoted by $(\Omega, {\cal F}, \p, \F)$
with real world probability measure $\p$ and (discrete time)
filtration $\F=({\cal F}_t)_{t \in \Delta \N_0}$. 

~

We assume that the ZCBs 
exist at all time points $t \in \Delta \N_0$ for all maturity
dates $T=t+m$ with
times to maturity $m\in \Delta \N$. Thus, we can consider
the discrete time yield curves
\begin{equation*}
\mathbf{Y}_t=( Y(t,t+m))'_{m \in \Delta \N}
%\qquad \text{ and } \qquad \mathbf{Y}^{abs}_t=( Y^{abs}(t,t+m))_{m \in \Delta \N}
\end{equation*}
for all time points $t\in \Delta \N_0$. Assume that $(\mathbf{Y}_t)_{t
\in \Delta \N_0}$ is $\F$-adapted, that is, 
$(\mathbf{Y}_s)_{s\le t}$ 
is observable at time $t$ and this information is contained
in the $\sigma$-field
${\cal F}_t$. Our aim is (as described above) to model and predict
$(\mathbf{Y}_t)_{t \in \Delta \N_0}$. We assume that there exists
an equivalent martingale measure $\p^\ast \sim \p$ for the bank
account numeraire discount $(B_t^{-1})_{t\in \Delta \N_0}$
and, in a first step,
we describe $(\mathbf{Y}_t)_{t \in \Delta \N_0}$ 
%and $(\mathbf{Y}^{abs}_t)_{t \in \Delta \N_0}$, respectively,
directly under this equivalent martingale measure $\p^\ast$. Notice here that the bank account
numeraire is actually a discrete 
time roll-over portfolio, as will be seen in the next section.

~

{\bf Remark.}
The assumption that the yield curve is given at any moment $ t \in \Delta \N_0 $ for sufficiently many maturities is a very strong one. In practice the yield curve is inter- and extrapolated every day from quite different traded quantities like coupon bearing bonds, swap rates, etc.
This inter- and extrapolation allows for a lot of freedom, often 
parametric families are used, e.g.~the Nelson-Siegel \cite{NelsonSiegel}
or the Svensson \cite{Svensson1, Svensson2}
family, but also non-parametric approaches such as 
splines are applied (see Filipovi\'c \cite{Damir}).

~

\section{Stochastic yield curve modeling and no-arbitrage}\label{modelling_na}

Assume the initial yield curve 
$\mathbf{Y}_0=(Y(0,m))_{m\in \Delta \N}$ 
at time $t=0$ is given.
For $t,m\in \Delta\N$ we make the following model assumptions: 
assume there exist deterministic functions
$\alpha_\Delta(\cdot, \cdot, \cdot)$ and  
$\mathbf{v}_\Delta(\cdot, \cdot, \cdot)$
such that the yield curve has the following stochastic
representation
\begin{eqnarray}
m~Y(t,t+m)&=&(m+\Delta)~Y(t-\Delta,t+m)-\Delta ~Y(t-\Delta,t)
\label{yield curve dynamics}
\\&&
+~{\alpha}_\Delta(t,m,(\mathbf{Y}_s)_{s\le t-\Delta})
+\mathbf{v}_\Delta(t,m,(\mathbf{Y}_s)_{s\le t-\Delta})~
\boldsymbol{\varepsilon}^\ast_t,\nonumber
\end{eqnarray}
where the innovations $\boldsymbol{\varepsilon}^\ast_t$
are ${\cal F}_t$-measurable and independent of ${\cal F}_{t-\Delta}$
under $\p^\ast$. In general, the innovations $\boldsymbol{\varepsilon}^\ast_t$
are multivariate
random vectors and the last product in \eqref{yield curve dynamics}
needs to be understood in the inner product sense.

~

{\bf Remark.}
The first two terms on the right-hand side of \eqref{yield curve dynamics}
will exactly correspond to the no-arbitrage condition in 
a deterministic interest rate model (see (2.2) in Filipovi\'c \cite{Damir}).
The fourth term on the right-hand side of \eqref{yield curve dynamics}
described by $\mathbf{v}_\Delta
(t,m,(\mathbf{Y}_s)_{s\le t-\Delta})
\boldsymbol{\varepsilon}^\ast_t$
adds the stochastic part to the future yield curve
development. Finally, the third term 
${\alpha}_\Delta(t,m,(\mathbf{Y}_s)_{s\le t-\Delta})$
will be recognized as a Heath-Jarrow-Morton \cite{HJM}
(HJM) term that makes the stochastic model free of arbitrage. This term is going to be analyzed in detail in Lemma \ref{HJM yield condition}
below. This approach allows us to separate conceptually the task of estimating volatilities, i.e.~estimating $ v_{\Delta} $, and estimating the market price of risk, i.e.~the difference of $ \p $ and $ \p^\ast $.

~

Assumption \eqref{yield curve dynamics} 
implies for the price of the ZCB
at time $t$ with time to maturity $m$ 
\begin{equation*}
P(t,t+m)=
\frac{P(t-\Delta,t+m)}{P(t-\Delta,t)}~
\exp \left\{
-{\alpha}_\Delta(t,m,(\mathbf{Y}_s)_{s\le t-\Delta})-
\mathbf{v}_\Delta(t,m,(\mathbf{Y}_s)_{s\le t-\Delta})~
\boldsymbol{\varepsilon}^\ast_t  \right\}.
\end{equation*}
In order to determine the HJM term
${\alpha}_\Delta(t,m,(\mathbf{Y}_s)_{s\le t-\Delta})$
we define the discrete time bank account value for an initial
investment of 1 as follows:
$B_0=1$ and for $t\in \Delta \N$
\begin{equation*}
B_t = \prod_{s=0}^{t/\Delta-1} P(\Delta s, \Delta (s+1))^{-1}
= \exp \left\{\Delta \sum_{s=0}^{t/\Delta-1} 
Y(\Delta s, \Delta (s+1))\right\}~>~0.
\end{equation*}
The process $\mathbf{B}=(B_t)_{t\in \Delta \N_0}$ considers the 
roll over of an initial
investment 1 into the (discrete time) bank account with grid 
size $\Delta$.
Note that $\mathbf{B}$ is previsible, i.e.~$B_t$
is ${\cal F}_{t-\Delta}$-measurable for all $t\in \Delta \N$.

~

Absence of arbitrage is now expressed in terms of the following 
$(\p^\ast,\F)$-martingale property (under the assumption that all the 
conditional
expectations exist). We require for all $t,m\in \Delta \N$ 
\begin{equation}\label{no arbitrage condition}
\E^\ast \left[\left.B_{t}^{-1}~ P(t,t+m)
\right|{\cal F}_{t-\Delta} \right]~\stackrel{!}{=}~
B_{t-\Delta}^{-1}~ P(t-\Delta,t+m).
\end{equation}
The necessity of such a martingale property is due to the fundamental theorem
of asset pricing (FTAP) derived in Delbaen-Schachermayer
\cite{DS}.
For notational convenience we set $\E^\ast_t\left[\cdot\right]
= \E^\ast\left[\left.\cdot \right|{\cal F}_t\right]$ for $t\in \Delta \N_0$.
The no-arbitrage condition \eqref{no arbitrage condition} immediately provides the following lemma.

\begin{lemma}\label{HJM yield condition}
Under the above assumptions the absence of arbitrage
condition \eqref{no arbitrage condition} 
implies
\begin{equation*}
{\alpha}_\Delta(t,m,(\mathbf{Y}_s)_{s\le t-\Delta})
=\log~
\E^\ast_{t-\Delta} \left[
\exp \left\{-
\mathbf{v}_\Delta(t,m,(\mathbf{Y}_s)_{s\le t-\Delta})~
\boldsymbol{\varepsilon}^\ast_t  \right\} \right].
\end{equation*}
\end{lemma}
This solves item (i) of the aim and scope list.

~

{\Beweis
{\bf Proof of Lemma \ref{HJM yield condition}.}
We rewrite  \eqref{no
arbitrage condition} as follows
(where we use assumption \eqref{yield curve dynamics}  
of the yield curve development and the appropriate 
measurability properties)
\begin{eqnarray*}
&&\hspace{-.51cm}
\exp \left\{- \Delta~
Y( t-\Delta, t)\right\}~
\E^\ast_{t-\Delta} \left[ P(t,t+m) \right]
~=~
P( t-\Delta, t)~
\E^\ast_{t-\Delta} \left[ P(t,t+m) \right]
\\
&&=~P(t-\Delta,t+m)~\exp \left\{-
{\alpha}_\Delta(t,m,(\mathbf{Y}_s)_{s\le t-\Delta})
\right\}
%\\&&~\times~
\E^\ast_{t-\Delta} \left[
\exp \left\{-
\mathbf{v}_\Delta(t,m,(\mathbf{Y}_s)_{s\le t-\Delta})~
\boldsymbol{\varepsilon}^\ast_t  \right\} \right]\\
&&\stackrel{!}{=}~P(t-\Delta,t+m).
\end{eqnarray*}
Solving this requirement proves the claim of Lemma \ref{HJM yield condition}.
\EndProof}

\section{Modeling aspects and calibration}

We need to discuss the choices 
$\mathbf{v}_\Delta(t,m,(\mathbf{Y}_s)_{s\le t-\Delta})$ and
$\boldsymbol{\varepsilon}^\ast_t$ as well as the description
of the equivalent martingale measure $\p^\ast\sim \p$.
Then, the model and the prediction
is fully specified through Lemma \ref{HJM yield
condition}.

\subsection{Data and explicit model choice}
Assume we would like to study a finite set ${\cal M}\subset \Delta \N$ of 
times to maturity.
We specify below necessary properties of ${\cal M}$ for yield curve prediction.
For these times to maturity choices we define for $t\in \Delta \N$
\begin{equation*}
\mathbf{Y}_{t, +}
=\left(Y(t,t+m)\right)_{m\in {\cal M}}'
\qquad \text{ and } 
\qquad 
\mathbf{Y}_{t, -}
=\left(Y(t-\Delta,t+m)\right)_{m\in {\cal M}}',
\end{equation*}
that is, in contrast to $\mathbf{Y}_{t}$
the random vectors $\mathbf{Y}_{t, +}$ 
and $\mathbf{Y}_{t, -}$ only
consider the times to maturity $m$ and $m+\Delta$
for $m\in {\cal M}$. 
Note that 
$\mathbf{Y}_{t, -}$ is ${\cal F}_{t-\Delta}$-measurable
and $\mathbf{Y}_{t, +}$ is ${\cal F}_{t}$-measurable.
Our aim is to model the change from $\mathbf{Y}_{t, -}$ to
 $\mathbf{Y}_{t, +}$. In view of \eqref{yield curve dynamics}
we define the vector
\begin{equation*}
\boldsymbol{\Upsilon}_t
~=~(\Upsilon_{t,m})'_{m\in {\cal M}}
~=~\left(m~Y(t,t+m)-(m+\Delta)~Y(t-\Delta,t+m)\right)_{m\in {\cal M}}'.
\end{equation*}
We set the dimension $d=|{\cal M}|$. For 
$\boldsymbol{\varepsilon}^\ast_t|_{{\cal F}_{t-\Delta}}$
we then choose
a $d$-dimensional standard Gaussian distribution with
independent components under the equivalent
martingale measure $\p^\ast$. 

~

{\bf Remark.}
We are aware that the choice of multivariate Gaussian innovations 
$\boldsymbol{\varepsilon}^\ast_t$ is only a first step towards more 
realistic innovation processes. However, we believe that already in this
model, with suitably chosen estimations of the instantaneous covariance 
structure, the results are quite convincing -- additionally chosen jump 
structures might even improve the situation. The independence assumption 
with respect to the martingale measure is an additional strong assumption 
which could be weakened.

~

Thus, we re-scale 
the volatility term with the grid size $\Delta$ and assume that at time $t$
it only
depends on the last observation $\mathbf{Y}_{t,-}$: define 
$\mathbf{v}_\Delta(\cdot, \cdot, \cdot)$ by
\begin{equation*}
\mathbf{v}_\Delta(t,m,(\mathbf{Y}_s)_{s\le t-\Delta})
=\sqrt{\Delta}~\boldsymbol{\sigma}(t,m,\mathbf{Y}_{ t,-}),
\end{equation*}
where the function $\boldsymbol{\sigma}(\cdot, \cdot, \cdot)$
does not depend on the grid size $\Delta$.
Lemma \ref{HJM yield condition}
implies for these choices for the HJM term
\begin{equation*}
{\alpha}_\Delta(t,m,(\mathbf{Y}_s)_{s\le t-\Delta})
=\log~
\E^\ast_{t-\Delta} \left[
\exp \left\{-\sqrt{\Delta}~
\boldsymbol{\sigma}(t,m,\mathbf{Y}_{t,-})~
\boldsymbol{\varepsilon}^\ast_t  \right\} \right]
%\\&=& 
=
\frac{\Delta}{2}~
\left\| \boldsymbol{\sigma}(t,m,\mathbf{Y}_{ t,-})
\right\|^2
%~=~ \frac{\Delta}{2}~
%\sum_{i=1}^d
%\boldsymbol{\sigma}_i^2(t,m,\mathbf{Y}_{ t-\Delta})
.
\end{equation*}
From \eqref{yield curve dynamics}
we then obtain for $t\in \Delta \N$ and $m\in {\cal M}$
under $\p^\ast$
\begin{equation}
\Upsilon_{t,m}
\label{yield curve dynamics 2}
= \Delta\left[
-Y(t-\Delta,t)
+\frac{1}{2}
\left\| \boldsymbol{\sigma}(t,m,\mathbf{Y}_{t,-})
\right\|^2\right]
+\sqrt{\Delta}~\boldsymbol{\sigma}(t,m,\mathbf{Y}_{ t,-})~
\boldsymbol{\varepsilon}^\ast_t.
\end{equation}
Note that $(\boldsymbol{\Upsilon}_t)_{t\in \Delta \N}$ 
is a $d$-dimensional process, thus,
we need a $d$-dimensional Gaussian random vector
$\boldsymbol{\varepsilon}^\ast_t|_{{\cal F}_{t-\Delta}}$ for 
obtaining full rank and no singularities. 
Next, we specify explicitly the $d$-dimensional function
$\boldsymbol{\sigma}(\cdot, \cdot, \cdot)$. 
We proceed similar to
Ortega et al.~\cite{Ortega}, i.e.~we directly model volatilities
and return directions.
Assume that for every $\mathbf{y}\in \R^d$ there exists 
an invertible and linear map
\begin{equation}\label{invertible map}
\varsigma(\mathbf{y}):\R^d \to \R^d, \qquad
\boldsymbol{\lambda} \mapsto 
\varsigma(\mathbf{y})(\boldsymbol{\lambda}).
\end{equation}
In the sequel we identify the linear map $\varsigma(\mathbf{y})(\cdot)$
with the corresponding (invertible)
matrix $\varsigma(\mathbf{y})\in \R^{d\times d}$
which generates this linear map, i.e.~$\varsigma(\mathbf{y})(\boldsymbol{\lambda})=\varsigma(\mathbf{y})~\boldsymbol{\lambda}$.
In the next step, we choose vectors 
$\boldsymbol{\lambda}_1, \ldots, \boldsymbol{\lambda}_d \in \R^d$
and define the matrix
${\Lambda}=
\left[
\boldsymbol{\lambda}_1, \ldots, \boldsymbol{\lambda}_d\right]\in \R^{d\times
d}$. Moreover, for $\mathbf{y}\in \R^d$ we set
\begin{eqnarray*}
\Sigma_{{\Lambda}} (\mathbf{y})
&=& \varsigma(\mathbf{y})~
\Lambda ~\Lambda' ~\varsigma'(\mathbf{y})
\in \R^{d\times d}.
\end{eqnarray*}
Using vector form
we make the following model
specification for \eqref{yield curve dynamics 2}:

\begin{model ass} \label{Model Assumptions 1}
We choose the following model
for the yield curve at time $t\in \Delta \N$ with
time to maturity dates ${\cal M}$:
\begin{equation*}
\boldsymbol{\Upsilon}_t
= \Delta\left[-\mathbf{Y}(t-\Delta,t)+\frac{1}{2}
~{\rm sp}
(\Sigma_{{\Lambda}} 
({\mathbf{Y}_{t, -}}))
\right]+ \sqrt{\Delta}~
\varsigma({\mathbf{Y}_{t, -}})~{\Lambda} 
~ \boldsymbol{\varepsilon}_{t}^\ast,
\end{equation*}
with $\mathbf{Y}(t-\Delta,t)=({Y}(t-\Delta,t),\ldots, {Y}(t-\Delta,t))'
\in \R^d$ and ${\rm sp}(\Sigma_{{\Lambda}})$ 
denotes the $d$-dimensional vector that
contains the diagonal elements of the matrix $\Sigma_{{\Lambda}}
 \in \R^{d\times d}$.
\end{model ass}
For the  
$j$-th maturity $m_j \in {\cal M}$
 we have done the following choice
\begin{equation*}
\boldsymbol{\sigma}(t,m_j,\mathbf{Y}_{t,-})~
\boldsymbol{\varepsilon}^\ast_t
=\sum_{i=1}^d
\boldsymbol{\sigma}_i(t,m_j,\mathbf{Y}_{t,-})~
{\varepsilon}^\ast_{t,i}
=\sum_{i=1}^d
\left[\varsigma(\mathbf{Y}_{t, -})
~ \boldsymbol{\lambda}_i\right]_j~
{\varepsilon}^\ast_{t,i}.
\end{equation*}
The linear map $\varsigma(\cdot)$ describes the
{\it volatility scaling factors},
$\boldsymbol{\lambda}_1, \ldots, \boldsymbol{\lambda}_d \in \R^d$
specify the {\it return directions}, and the volatility choice
does not depend on the grid size $\Delta$.
Our aim is to calibrate these terms.

~

{\bf Remark.}\label{remark1}
The volatility scaling factors $\varsigma(\cdot)$ mimic how volatility for different maturities scales with the level of yield at this maturity. Several approaches have been discussed in the literature. The choice of a square-root dependence seems to be quite robust over different maturities and interest rate regimes, but for small rates -- as we face it for the Swiss currency CHF -- linear dependence seems to be a good choice, too, see choice \eqref{choice sigma}.

\begin{lemma} \label{Lemma 3.2}
Under Model Assumptions \ref{Model Assumptions 1}, the random 
vector
$\boldsymbol{\Upsilon}_t|_{{\cal F}_{t-\Delta}}$ has
a $d$-dimensional conditional Gaussian distribution
with the first two conditional moments given by
\begin{eqnarray*}
\E^\ast_{t-\Delta} \left[
\boldsymbol{\Upsilon}_t\right]&=&
 \Delta\left[-\mathbf{Y}(t-\Delta,t)+\frac{1}{2}
~{\rm sp}(\Sigma_{\Lambda} ({\mathbf{Y}_{t, -}}))
\right],\\
{\rm Cov}^\ast_{t-\Delta } \left(
\boldsymbol{\Upsilon}_t\right)&=&
\Delta ~
\Sigma_{\Lambda} 
({\mathbf{Y}_{t, -}}).
\end{eqnarray*}
\end{lemma}

\subsection{Calibration procedure}
In order to calibrate our model
we need to choose the volatility scaling factors  $\varsigma(\cdot)$
and we need to
specify the return directions
$\boldsymbol{\lambda}_1, \ldots, \boldsymbol{\lambda}_d \in \R^d$
which provide the matrix $\Lambda$. In fact we do not need to specify the direction $\boldsymbol{\lambda}_1, \ldots, \boldsymbol{\lambda}_d \in \R^d$
themselves, but rather $ \Sigma_{\Lambda} $, which we shall do in the sequel.
Assume we have observations 
$(\boldsymbol{\Upsilon}_t)_{t=\Delta,\ldots, \Delta K}$,
$(Y(t-\Delta,t))_{t=\Delta,\ldots \Delta(K+1)}$,
and $(\mathbf{Y}_{t,-})_{t=\Delta,\ldots, \Delta (K+1)}$. We 
use these observations to predict/approximate the
random vector $\boldsymbol{\Upsilon}_{\Delta (K+1)}$
at time $\Delta K$.
For $\mathbf{y}\in \R^d$ we define the matrices
\begin{eqnarray*}
C_{(K)} &=&\frac{1}{\sqrt{K}} \left(\left[
\varsigma(\mathbf{Y}_{\Delta k,-})^{-1}~
\boldsymbol{\Upsilon}_{\Delta k}\right]_j
\right)_{j=1,\ldots, d; ~k=1, \ldots, K}\in \R^{d\times K},\\
S_{(K)}(\mathbf{y})&=&
\varsigma(\mathbf{y})~
C_{(K)}~C'_{(K)}~\varsigma'(\mathbf{y})\in \R^{d\times d}.
\end{eqnarray*}
Choose $t=\Delta(K+1)$.
Note that $C_{(K)}$ is ${\cal F}_{t-\Delta}$-measurable.
For $\mathbf{x},\mathbf{y}\in \R^d$
we define the $d$-dimensional random vector
\begin{eqnarray}\label{model kappa}
\boldsymbol{\kappa}_t~=~\boldsymbol{\kappa}_t(\mathbf{x},\mathbf{y})
&=& - \Delta~\mathbf{x}+\frac{1}{2}~
{\rm sp}\left(S_{(K)}(\mathbf{y})\right)
+ \varsigma(\mathbf{y})~
C_{(K)}~ \mathbf{W}^\ast_{t},
\end{eqnarray}
with $\mathbf{W}^\ast_{t}$ is independent
of ${\cal F}_{t-\Delta}$, ${\cal F}_t$-measurable, independent
of $\boldsymbol{\varepsilon}^\ast_t$ and a $K$-dimensional
standard Gaussian random vector
with independent components under $\p^\ast$.
\begin{lemma} \label{Lemma 3.3} The random 
vector
$\boldsymbol{\kappa}_t|_{{\cal F}_{t-\Delta}}$ has
a $d$-dimensional  Gaussian distribution
with the first two conditional moments given by
\begin{eqnarray*}
\E^\ast_{t-\Delta} \left[
\boldsymbol{\kappa}_t\right]&=&
- \Delta~\mathbf{x}+\frac{1}{2}~
{\rm sp}\left(S_{(K)}(\mathbf{y})\right),\\
{\rm Cov}^\ast_{t-\Delta} \left(
\boldsymbol{\kappa}_t\right)&=&
S_{(K)}(\mathbf{y}).
\end{eqnarray*}
\end{lemma}
Our aim is to show that the matrix $S_{(K)}(\mathbf{y})$
is an appropriate estimator for 
$\Delta 
\Sigma_{\Lambda} 
(\mathbf{y})$ and then Lemmas \ref{Lemma 3.2} and \ref{Lemma 3.3}
say that $\boldsymbol{\kappa}_t$ is an appropriate stochastic
approximation to $\boldsymbol{\Upsilon}_t$, conditionally
given ${\cal F}_{t-\Delta}$.

~

{\bf Remark.} The random vector $\boldsymbol{\kappa}_t$ can be seen
as a filtered historical simulation where $\mathbf{W}^\ast_{t}$
re-simulates the $K$ observations which are appropriately 
historically scaled through the matrix $C_{(K)}$.

~

We calculate the expected value of $S_{(K)}(\mathbf{y})$ under $\p^\ast$.
Choose $\mathbf{z},\mathbf{y} \in \R^d$ and define the function
\begin{equation*}
f_\Lambda(\mathbf{z},\mathbf{y})
= \varsigma(\mathbf{y})^{-1}~
\left[-\mathbf{z}+\frac{1}{2}{\rm sp}\left(
\Sigma_{\Lambda} ({\mathbf{y}})
\right)
\right]\left[-\mathbf{z}+\frac{1}{2}{\rm sp}\left(
\Sigma_{\Lambda} ({\mathbf{y}})
\right)
\right]'\left(\varsigma(\mathbf{y})^{-1}\right)'.
\end{equation*}
Note that this function does {\it not} depend on the grid size $\Delta$.
Lemma \ref{Lemma 3.2} then implies
that
\begin{equation}\label{function f analysis}
f_\Lambda(\mathbf{Y}(t-\Delta,t),\mathbf{Y}_{t, -})~=~
 \Delta^{-2}~
\varsigma(\mathbf{Y}_{t, -})^{-1}~
\E^\ast_{t-\Delta} \left[
\boldsymbol{\Upsilon}_t\right]~\E^\ast_{t-\Delta} \left[
\boldsymbol{\Upsilon}_t\right]'
\left(\varsigma(\mathbf{Y}_{t, -})^{-1}\right)',
\end{equation}
where the left-hand side only depends on $\Delta$ through the fact
that the yield curve $\mathbf{Y}_{t-\Delta}$ is observed at time $t-\Delta$, however otherwise
it does not depend on $\Delta$ (as a scaling factor).

\begin{theo}\label{theorem unbiased}
Under Model Assumptions \ref{Model Assumptions 1} we obtain
for all $K\in \N$ and $\mathbf{y}\in \R^d$
\begin{equation*}
\E^\ast_{0}\left[S_{(K)}(\mathbf{y})\right]
=\Delta~
\Sigma_{\Lambda} 
(\mathbf{y})+\Delta^2~
\varsigma(\mathbf{y}) \left(\frac{1}{K}\sum_{k=1}^K
\E^\ast_{0}\left[
f_\Lambda(\mathbf{Y}(\Delta(k-1),\Delta k),\mathbf{Y}_{\Delta k, -})\right]
\right)
 \varsigma(\mathbf{y})'.
\end{equation*}
\end{theo}

{\bf Interpretation.}
Using $S_{(K)}(\mathbf{y})$ as estimator for $\Delta
\Sigma_{\Lambda}  (\mathbf{y})$ provides, under $\p^\ast_0$, a bias given by
\begin{equation*}
\Delta^2~
\varsigma(\mathbf{y}) \left(\frac{1}{K}\sum_{k=1}^K
\E^\ast_{0}\left[
f_\Lambda(\mathbf{Y}(\Delta(k-1),\Delta k),\mathbf{Y}_{\Delta k, -})\right]
\right)
 \varsigma(\mathbf{y})'.
\end{equation*}
If we choose $t=\Delta K$ fixed and assume that the
term in the bracket is uniformly bounded for $\Delta \to 0$ then we see that
\begin{equation}\label{asymptotic result}
\E^\ast_{0}\left[S_{(K)}(\mathbf{y})\right]
=\Delta~
\Sigma_{\Lambda} 
(\mathbf{y})+\Delta^2 ~O(\mathbf{1}),\qquad \text{ for }\Delta\to 0.
\end{equation}
That is, for small grid size $\Delta$ the second term 
should become negligible.  

~

The only term that still needs to be chosen is the
invertible and linear map $\varsigma(\mathbf{y})$, i.e.~the 
volatility scaling factors.
For $\vartheta \ge 0$ we define that function
\begin{equation}\label{h function}
h:\R_+ \to \R_+, \qquad y \mapsto
h(y)=\vartheta^{-1/2}~y 1_{\{y \le \vartheta \}}+y^{1/2} 1_{\{y > \vartheta \}}.
\end{equation}
As already remarked in Subsection \ref{remark1} 
in the literature one often finds
the square-root scaling, however for small rates a linear scaling
can also be appropriate. For the Swiss currency CHF it turns out below
that the linear scaling is appropriate for a threshold
of $\vartheta=2.5\%$. In addition, we  define the
function $h(\cdot)$ as above to guarantee that the processes
do not explode for large volatilities and small grid sizes.

Assume that there exist constants $\sigma_j>0$, then
we set for $\mathbf{y}=(y_1,\ldots, y_d)\in \R^d$
\begin{eqnarray}
\varsigma(\mathbf{y})&=& {\rm diag}(\sigma_1 h(y_1) ,\ldots, 
\sigma_d h(y_d))~
=~
 {\rm diag}(\sigma_1,\ldots, 
\sigma_d)~
 {\rm diag}(h(y_1),\ldots, h(y_d)).
\nonumber
\end{eqnarray}
Basically, volatility is scaled according to the actual
observation $\mathbf{y}$. This choice implies
\begin{eqnarray*}
\varsigma(\mathbf{y})~C_{(K)} &=&\frac{1}{\sqrt{K}}~\varsigma(\mathbf{y})~ \left(\left[
\varsigma(\mathbf{Y}_{\Delta k,-})^{-1}~
\boldsymbol{\Upsilon}_{\Delta k}\right]_j
\right)_{j=1,\ldots, d; ~k=1, \ldots, K}\\
&=&\frac{1}{\sqrt{K}}~ {\rm diag}(h(y_1),\ldots, 
h(y_d))~ \left(\left[
 {\rm diag}(h(\mathbf{Y}_{\Delta k,-}))^{-1}~
\boldsymbol{\Upsilon}_{\Delta k}\right]_j
\right)_{j=1,\ldots, d; ~k=1, \ldots, K},
\end{eqnarray*}
thus, the constants $\sigma_j>0$ do not need to be estimated because they are
already (implicitly) contained in the observations and, hence, in $\Lambda$.
Therefore, we set them to 1 and we choose
\begin{eqnarray}
\varsigma(\mathbf{y})&=& 
 {\rm diag}(h(y_1),\ldots, h(y_d)).
\label{choice sigma}
\end{eqnarray}
These assumptions now allow to directly
analyze the bias term given in \eqref{asymptotic result}.
Therefore, we need to evaluate the function $f_\Lambda$
in Theorem \ref{theorem unbiased}.
However, to this end we would need to know $\Sigma_\Lambda$, 
i.e.~we obtain from Theorem \ref{theorem unbiased} an implicit
solution (quadratic form) that can be solved for $\Sigma_\Lambda$.
We set $\mathbf{y}=\mathbf{1}$ and then obtain
from Theorem \ref{theorem unbiased}
\begin{eqnarray*}
{\Delta}^{-1}~
\E^\ast_{0}\left[S_{(K)}(\mathbf{1})\right]
&=&
\Sigma_{\Lambda} 
(\mathbf{1})+\Delta
 \left(\frac{1}{K}\sum_{k=1}^K
\E^\ast_{0}\left[
f_\Lambda(\mathbf{Y}(\Delta(k-1),\Delta k),\mathbf{Y}_{\Delta k, -})\right]
\right).
\end{eqnarray*}
Note that $\Sigma_{\Lambda} (\mathbf{y})
=\varsigma(\mathbf{y})\Lambda\Lambda'\varsigma(\mathbf{y})$, thus
under \eqref{choice sigma} its  elements are
given by $h(y_i)h(y_j)s_{ij}$, $i,j=1,\ldots, d$, where we have defined
$\Lambda\Lambda'=(s_{ij})_{i,j=1,\ldots, d}$.
Let us first concentrate on the diagonal elements, i.e.~$i=j$, and
assume that time to maturity $m_i$ corresponds to index $i$.
\begin{eqnarray*}
\Delta^{-1}\left(
\E^\ast_{0}\left[{S_{(K)}(\mathbf{1})}\right]\right)_{ii}
&=&s_{ii}+
\frac{\Delta}{K}\sum_{k=1}^K
\Bigg(
\E^\ast_{0}\left[
\left(\frac{Y(\Delta(k-1),\Delta k)}
{h(Y(\Delta(k-1),\Delta k+m_i))}\right)^2\right]
\\
&&
+~\frac{1}{4}
\E^\ast_{0}\left[h(Y(\Delta(k-1),\Delta k+m_i))^2\right]s^2_{ii}
-\E^\ast_{0}\left[Y(\Delta(k-1),\Delta k)\right]s_{ii}\Bigg).
\end{eqnarray*}
This is a quadratic equation that can be solved for $s_{ii}$.
Define 
\begin{eqnarray}
\label{a_i calculation}
a_i&=&\frac{\Delta}{4K}\sum_{k=1}^K
\E^\ast_{0}\left[h(Y(\Delta(k-1),\Delta k+m_i))^2\right],\\
\label{b calculation}
b&=&
1-
\frac{\Delta}{K}\sum_{k=1}^K\E^\ast_{0}\left[Y(\Delta(k-1),\Delta k)\right],
\\
\label{c_i calculation}
c_i&=&
-\Delta^{-1}\left(
\E^\ast_{0}\left[{S_{(K)}(\mathbf{1})}\right]\right)_{ii}
+\frac{\Delta}{K}\sum_{k=1}^K
\E^\ast_{0}\left[
\left(\frac{Y(\Delta(k-1),\Delta k)}
{h(Y(\Delta(k-1),\Delta k+m_i))}\right)^2\right],
\end{eqnarray}
then we have $a_i s_{ii}^2+bs_{ii}+c_i=0$ which provides the solution
\begin{equation}
\label{s_ii calculation}
s_{ii}= \frac{-b + \sqrt{b^2-4a_ic_i}}{2a_i}.
\end{equation}
Thus, the bias terms of the diagonal elements are given by
\begin{equation*}
\beta_{ii}=
\Delta^{-1}\left(
\E^\ast_{0}\left[{S_{(K)}(\mathbf{1})}\right]\right)_{ii}
-s_{ii},
\end{equation*}
which we are going to analyze below for the different maturities $m_i\in
{\cal M}$. For the off-diagonals $i\neq j$ and the corresponding
maturities $m_i$ and $m_j$ we obtain
{\small
\begin{eqnarray}
\nonumber
\Delta^{-1}\left(
\E^\ast_{0}\left[{S_{(K)}(\mathbf{1})}\right]\right)_{ij}
&=&s_{ij}+
\frac{\Delta}{K}\sum_{k=1}^K
\Bigg(
\E^\ast_{0}\left[
\frac{Y(\Delta(k-1),\Delta k)}
{h(Y(\Delta(k-1),\Delta k+m_i))}
\frac{Y(\Delta(k-1),\Delta k)}
{h(Y(\Delta(k-1),\Delta k+m_j))}
\right]
\\\nonumber
&&\qquad
+\frac{1}{4}
\E^\ast_{0}\left[h(Y(\Delta(k-1),\Delta k+m_i))
h(Y(\Delta(k-1),\Delta k+m_j))\right]
s_{ii}s_{jj}\\&&\label{s_ij calculation}
\qquad
-\frac{1}{2}\E^\ast_{0}\left[Y(\Delta(k-1),\Delta k)
\frac{Y(\Delta(k-1),\Delta k+m_i)}
{h(Y(\Delta(k-1),\Delta k+m_j))}\right]s_{ii}
\\&&\nonumber
\qquad
-\frac{1}{2}\E^\ast_{0}\left[Y(\Delta(k-1),\Delta k)
\frac{Y(\Delta(k-1),\Delta k+m_j)}
{h(Y(\Delta(k-1),\Delta k+m_i))}
\right]s_{jj}
\Bigg).
\end{eqnarray}}
This can easily be solved for $s_{ij}$ for given $s_{ii}$ and $s_{jj}$.

\section{Calibration to real data}
\subsection{Calibration}

For the time-being we assume that
$\p=\p^\ast$, i.e.~we assume that the market price of risk is identical
equal to 0. This simplifies the calibration and as a consequence we can
directly work on the observed data. The choice of the drift
term will be discussed below. 

The first difficulty is the choice of the data. The reason therefore
is that risk-free ZCBs do {\it not} exist and, thus, the risk-free
yield curve needs to be estimated from data that has different spreads
such as a credit spread, a liquidity spread, a long-term premium, etc.

\begin{figure}[htp]
\vspace{-1cm}
\begin{center}
\includegraphics[width=\linewidth]{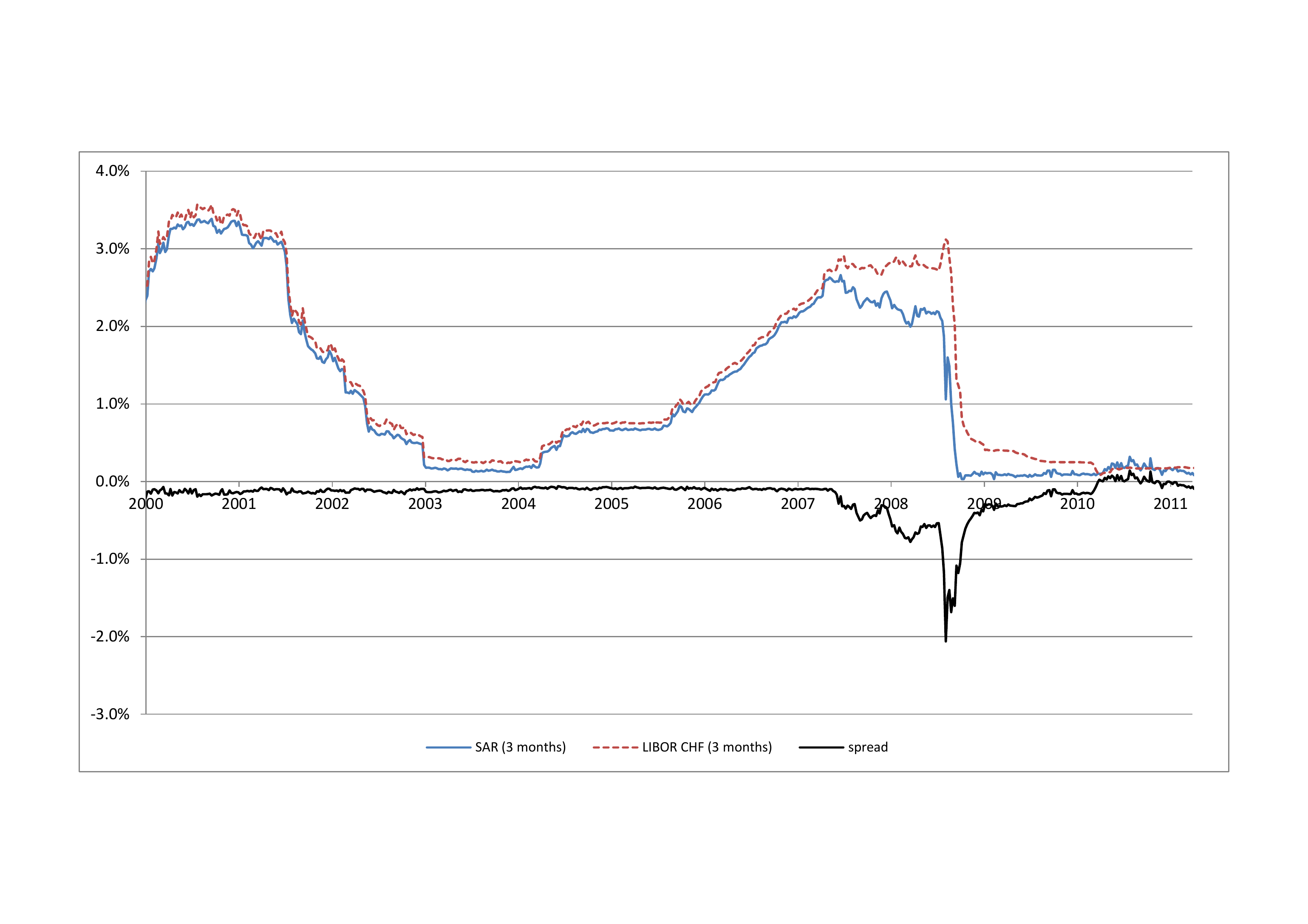}
\end{center}
\vspace{-2cm} 
\caption{Yield curve time series 3 Months SAR and 3 Months CHF LIBOR
from 01/2000 until 05/2011. The spread gives the difference between
these two time series.} \label{Figure 1}
%\end{figure}

%\begin{figure}[htp]
\vspace{-1cm}
\begin{center}
\includegraphics[width=\linewidth]{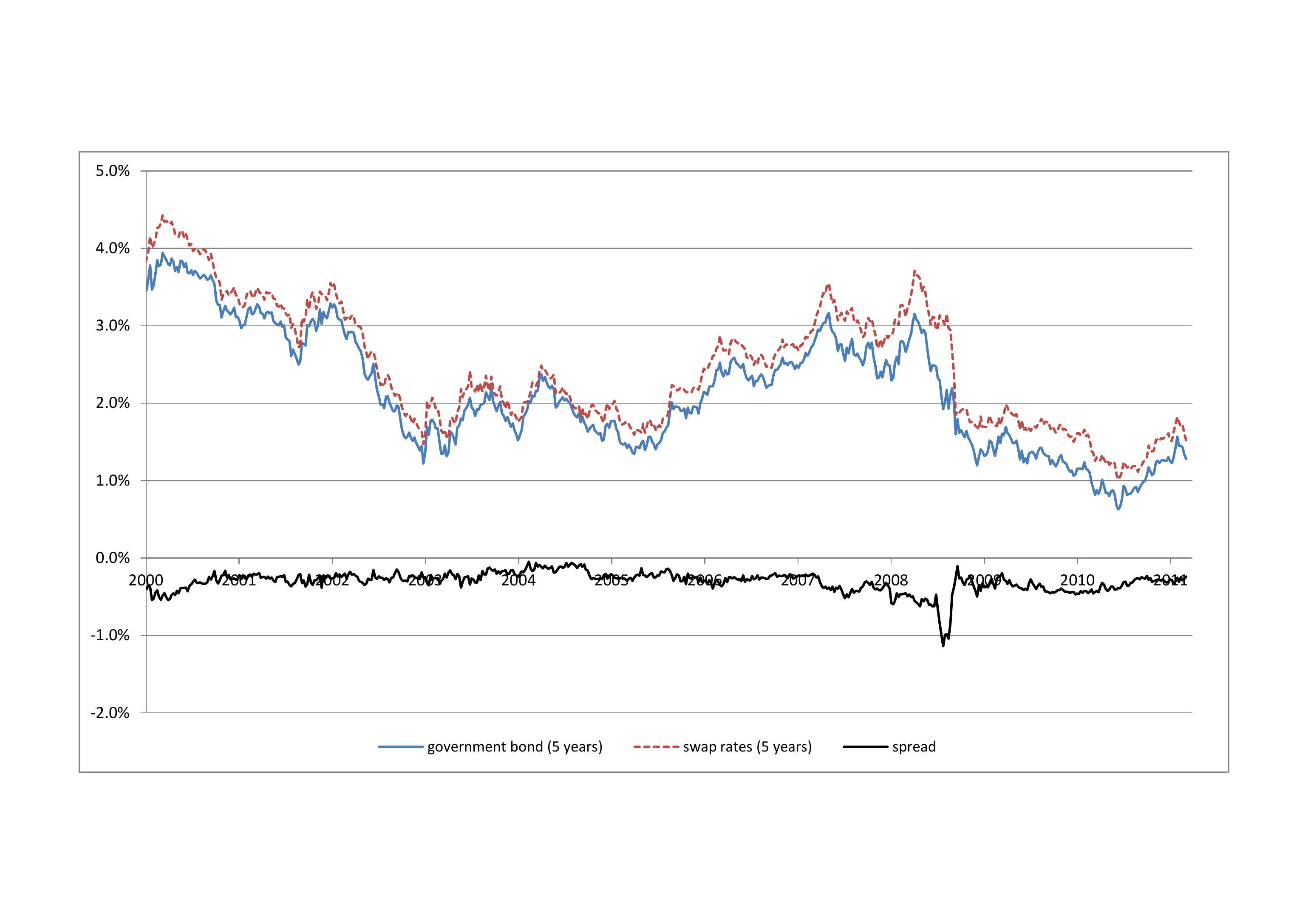}
\end{center}
\vspace{-2cm} 
\caption{Yield curve time series Swiss government bond and
CHF swap rate both for time to maturity $m=5$ years.
The spread gives the difference between
these two time series.} \label{Figure 2}
\end{figure}

We calibrate the model to the Swiss currency CHF.
For short times to maturity (below one year) one typically chooses
either the LIBOR (London InterBank Offered Rate)
or the SAR (Swiss Average Rate), see Jordan \cite{Jordan}, as (almost)
risk-free financial instruments. The LIBOR is the rate at which
highly-credit banks borrow and lend money at the inter-bank market.
The SAR is a rate determined by the Swiss National Bank at which
highly-credited institutions borrow and lend money with securization.
We display the yields of these two financial time series for instruments of a 
time to maturity of 3 months, see Figure \ref{Figure 1}. We see that
the SAR yield typically lies below the LIBOR yield
(due to securization). Therefore,
we consider the SAR to be less risky and we choose it  as approximation
to a risk-free financial instrument with short time to maturity.

For long times to maturity (above one year)
one either chooses government bonds 
(of sufficiently highly rated countries) or swap rates.
In Figure \ref{Figure 2} we give the time series of the Swiss government
bond and the CHF swap yields both for a time to maturity of 5 years.
We see that the rate of the Swiss government bond is below the swap rate 
(due to lower credit risk and maybe an illiquidity premium coming from a
high demand) and
therefore we choose Swiss government bonds as approximation to the
risk-free yield curve data for long times to maturity.

We mention that these short terms and long terms data are not completely
compatible which may give some difficulties in the calibration. We will
also see this in the correlation matrices below. 

\begin{figure}[ht]
\vspace{-1cm}
\begin{center}
\includegraphics[width=\linewidth]{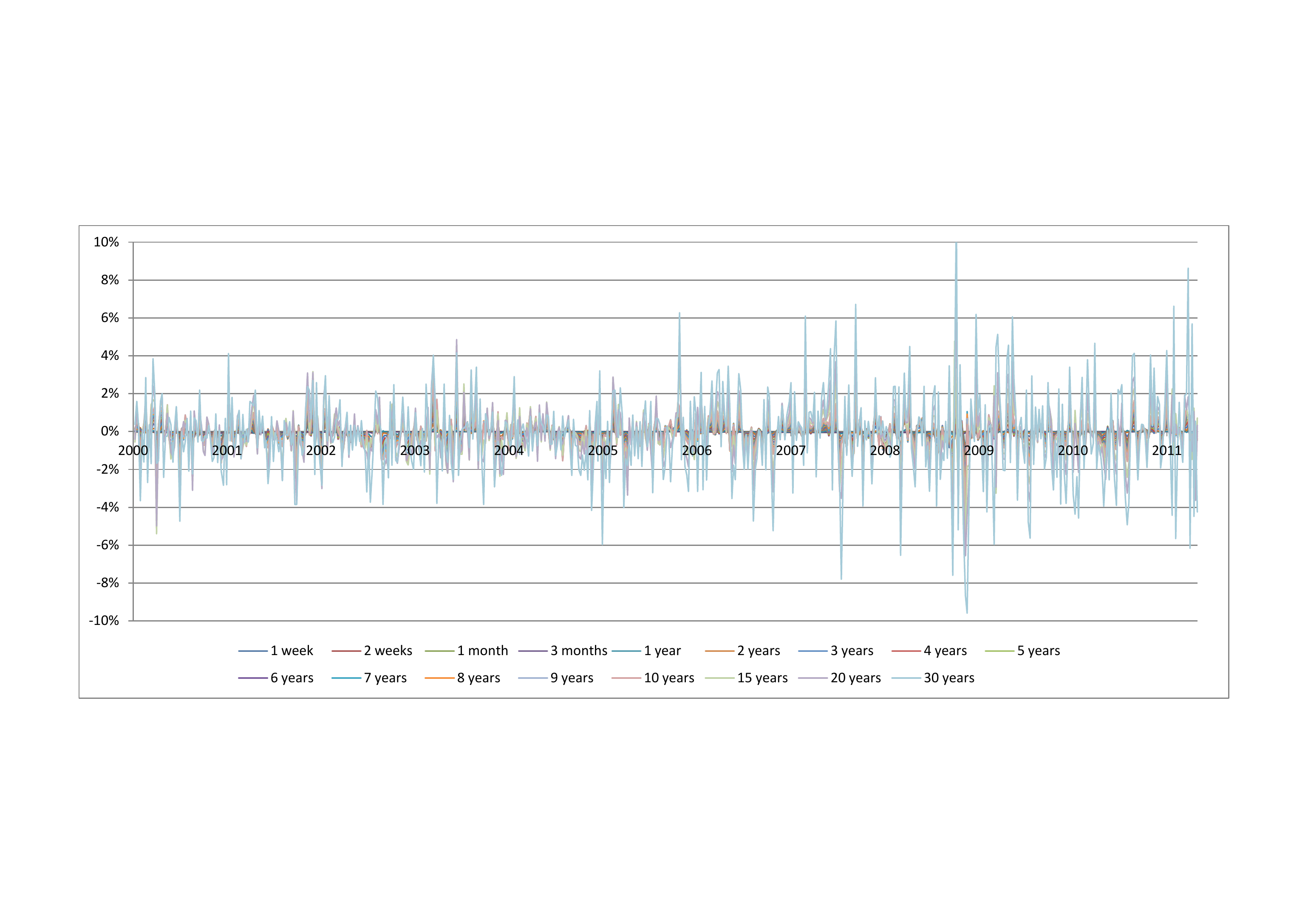}
\end{center}
\vspace{-3cm} 
\caption{Time series $\boldsymbol{\Upsilon}_t$ for
$t \in \{01/2000,\ldots, 05/2011\}$ on a weekly
grid $\Delta=1/52$.} \label{Figure 3}
%\end{figure}

%\begin{figure}[ht]
\vspace{-1cm}
\begin{center}
\includegraphics[width=\linewidth]{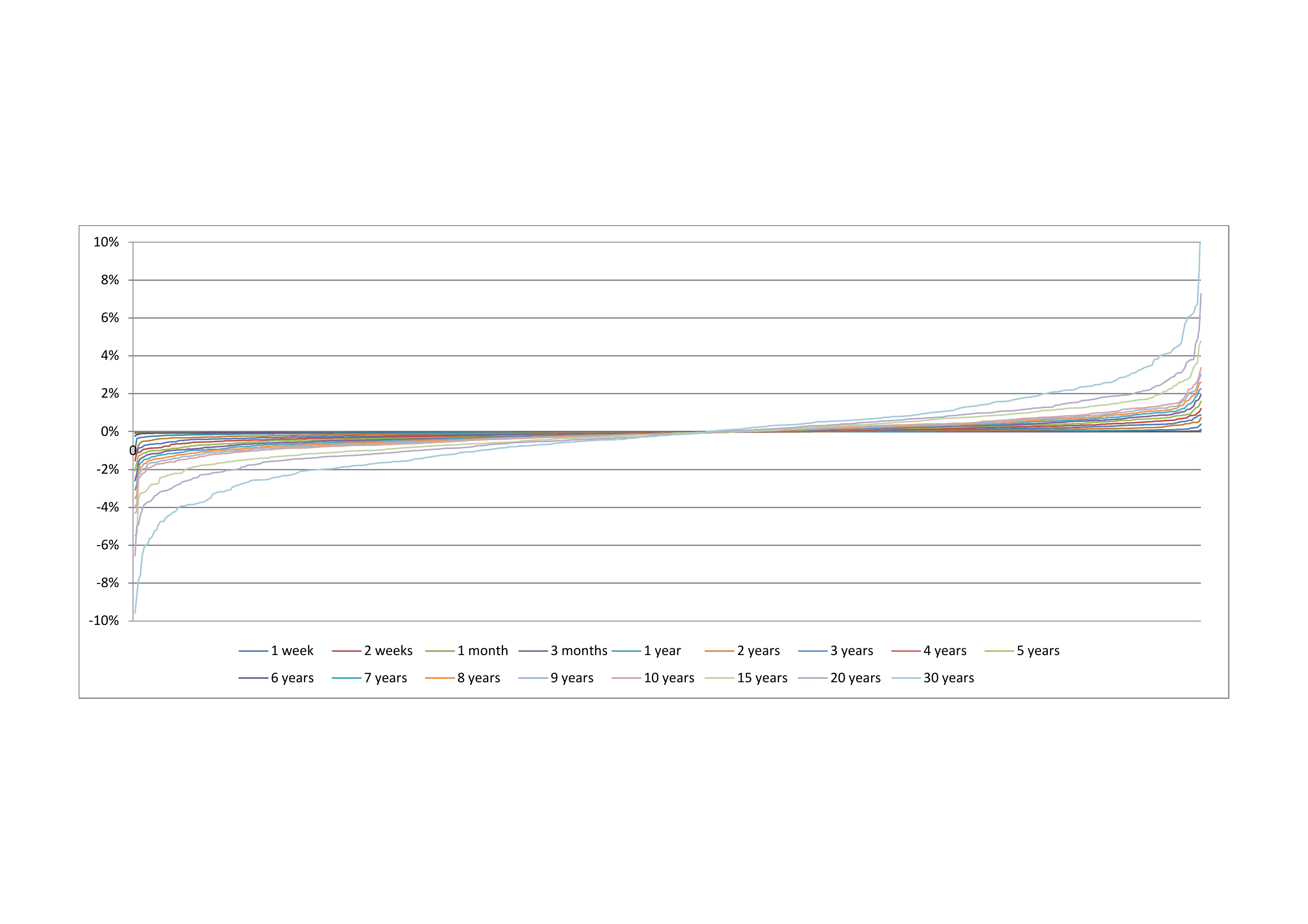}
\end{center}
\vspace{-3cm} 
\caption{Component-wise ordered time series obtained from
$\boldsymbol{\Upsilon}_{t}$ for
$t \in \{01/2000,\ldots, 05/2011\}$, i.e.~$\Upsilon_{(t),m}
\le \Upsilon_{(t+1),m}$ for all $t$ and $m\in {\cal M}$
on a weekly
grid $\Delta=1/52$.} \label{Figure 4}
\end{figure}

\begin{figure}[ht]
\vspace{-1cm}
\begin{center}
\includegraphics[width=\linewidth]{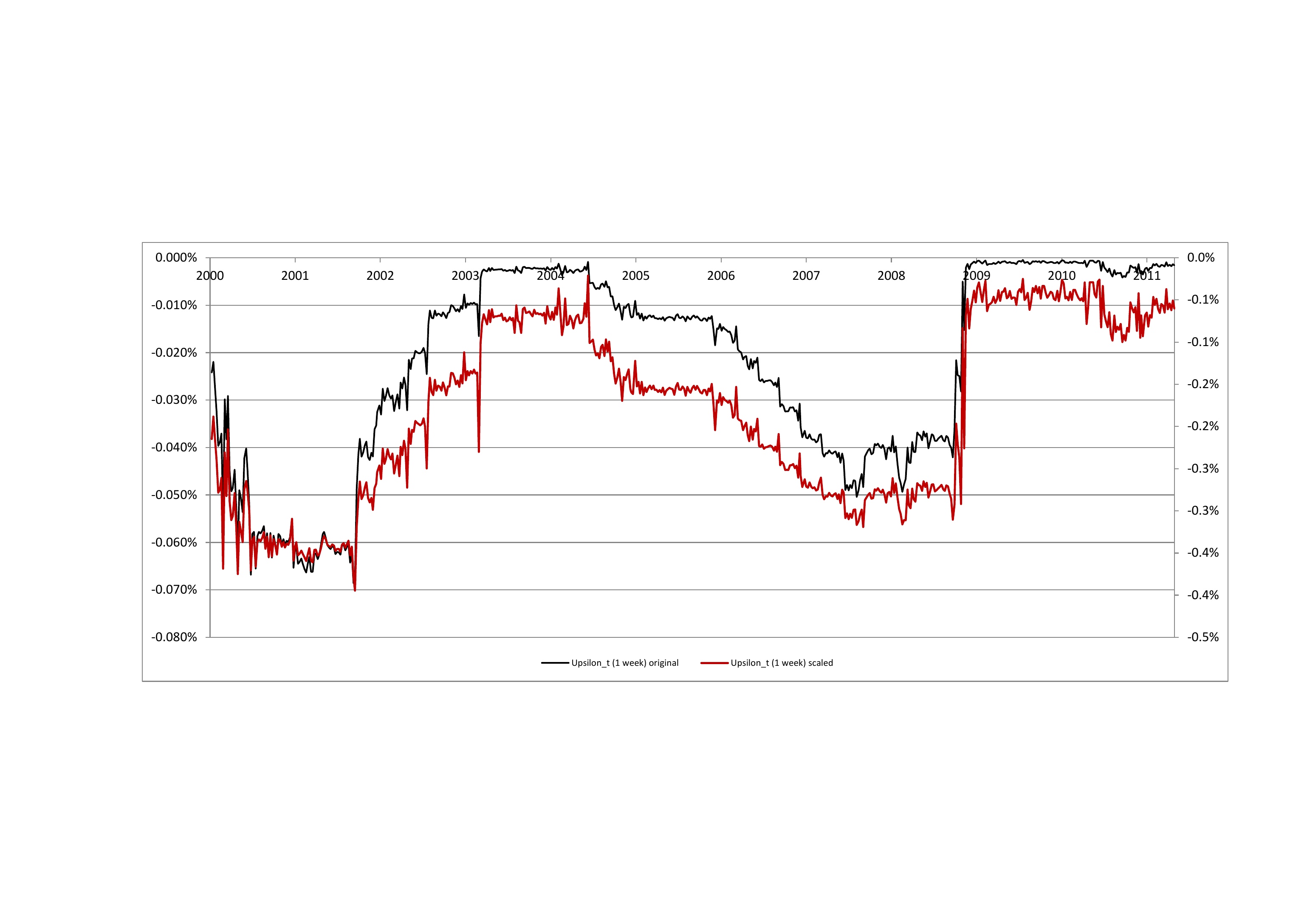}
\end{center}
\vspace{-3cm} 
\caption{Time series 
$\Upsilon_{t,m}$ and 
$[\sqrt{K} ~ C_{(K)}]_m=
\Upsilon_{t,m}/h(Y(t-\Delta, t+m))$ for
maturity $m=1/52$ and 
$t \in \{01/2000,\ldots, 05/2011\}$ on a weekly
grid $\Delta=1/52$.} \label{Figure 5}
%\end{figure}

%\begin{figure}[ht]
\vspace{-1cm}
\begin{center}
\includegraphics[width=\linewidth]{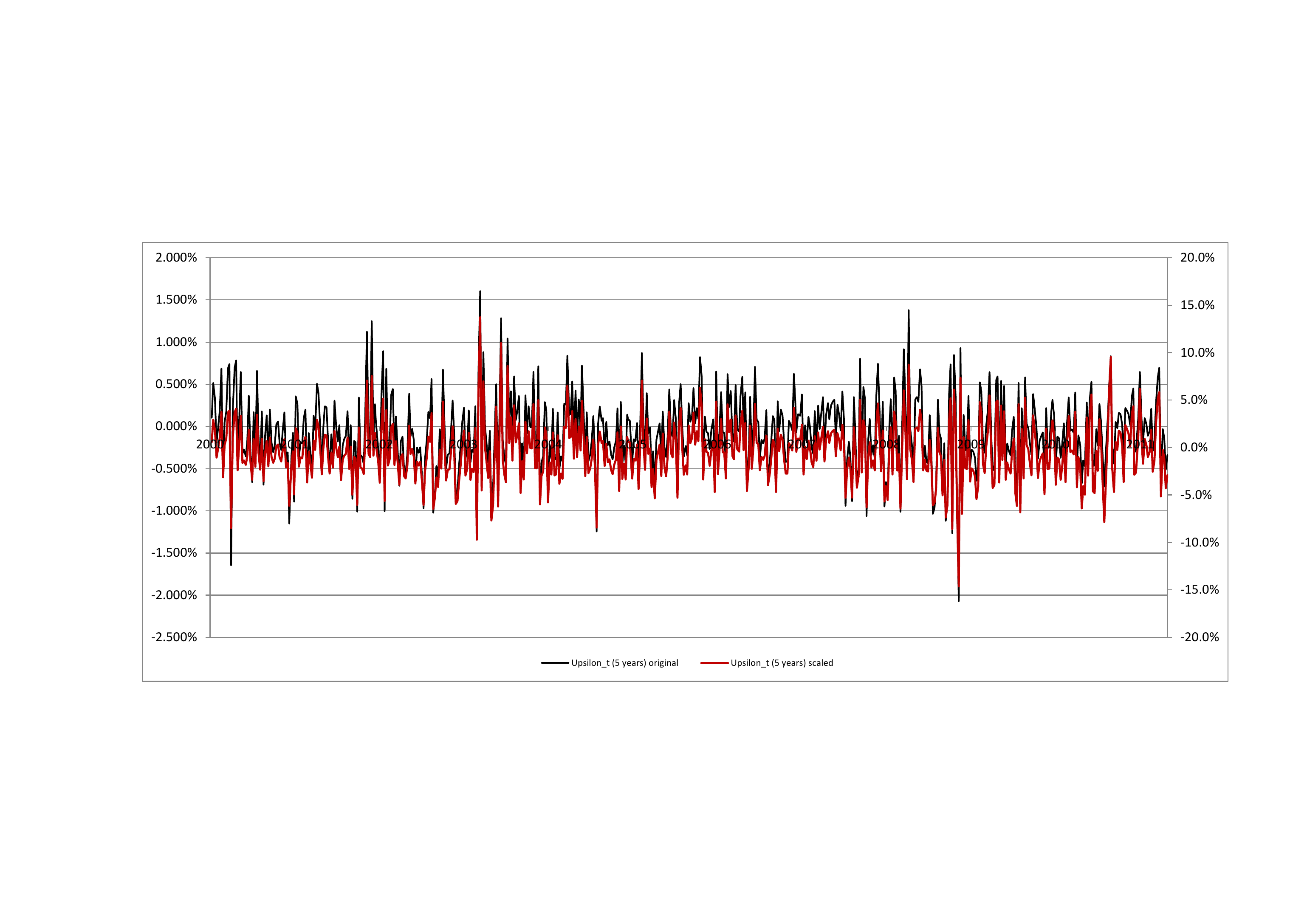}
\end{center}
\vspace{-3cm} 
\caption{Time series 
$\Upsilon_{t,m}$ and 
$[\sqrt{K} ~ C_{(K)}]_m=
\Upsilon_{t,m}/h(Y(t-\Delta, t+m))$ for
maturity $m=5$ and 
$t \in \{01/2000,\ldots, 05/2011\}$
on a weekly
grid $\Delta=1/52$.} \label{Figure 6}
\end{figure}

\begin{figure}[ht]
\vspace{-1cm}
\begin{center}
\includegraphics[width=\linewidth]{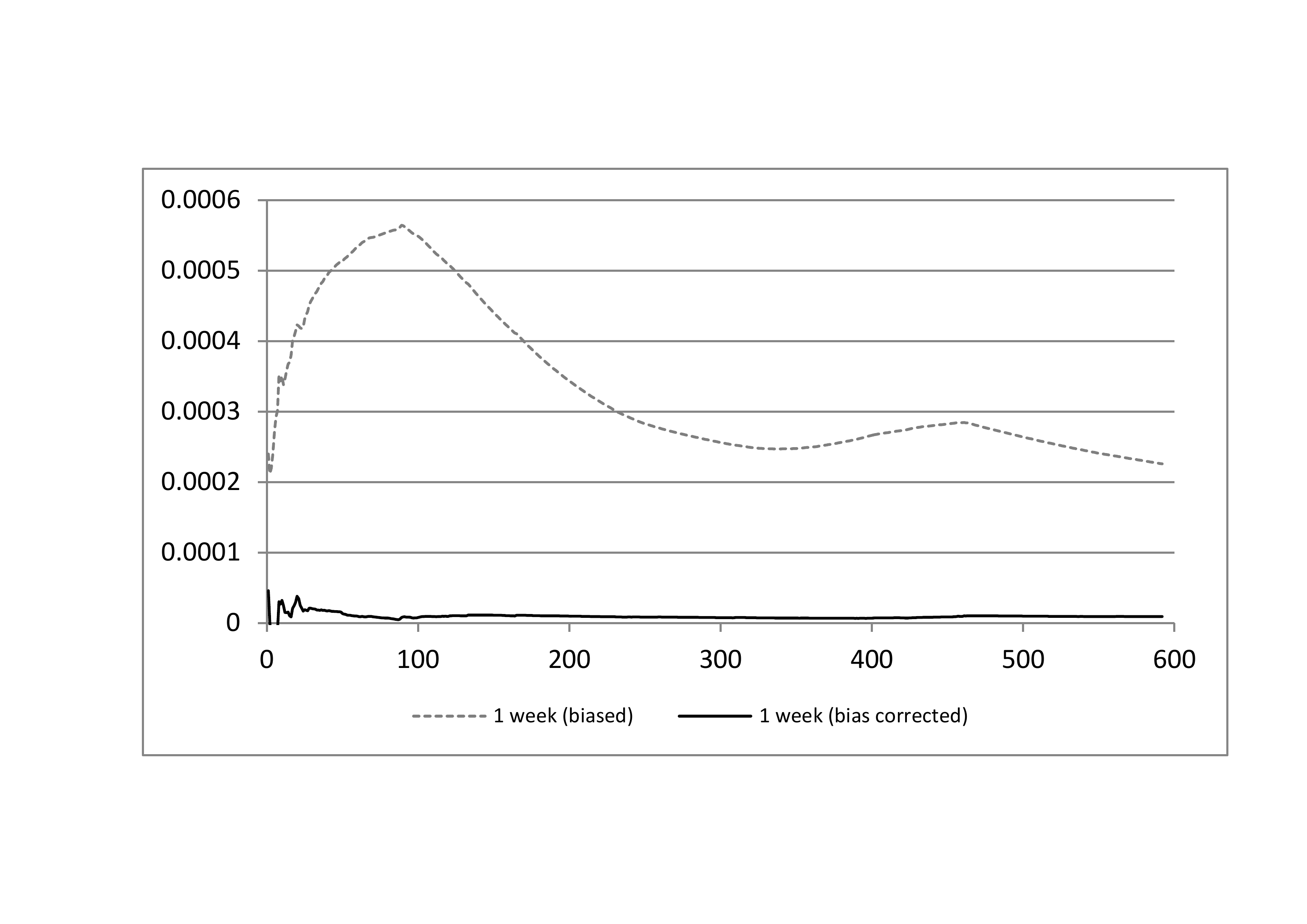}
\end{center}
\vspace{-2cm} 
\caption{Time series  $\widehat{s}_{ii}^{\rm bias}(K)$
and $\widehat{s}_{ii}(K)$, $K=1,\ldots, 600$,
 for maturity $m_i=1$ week and observations in
$ \{01/2000,\ldots, 05/2011\}$
on a weekly
grid $\Delta=1/52$.} \label{Figure 7}
%\end{figure}

%\begin{figure}[ht]
\vspace{-1cm}
\begin{center}
\includegraphics[width=\linewidth]{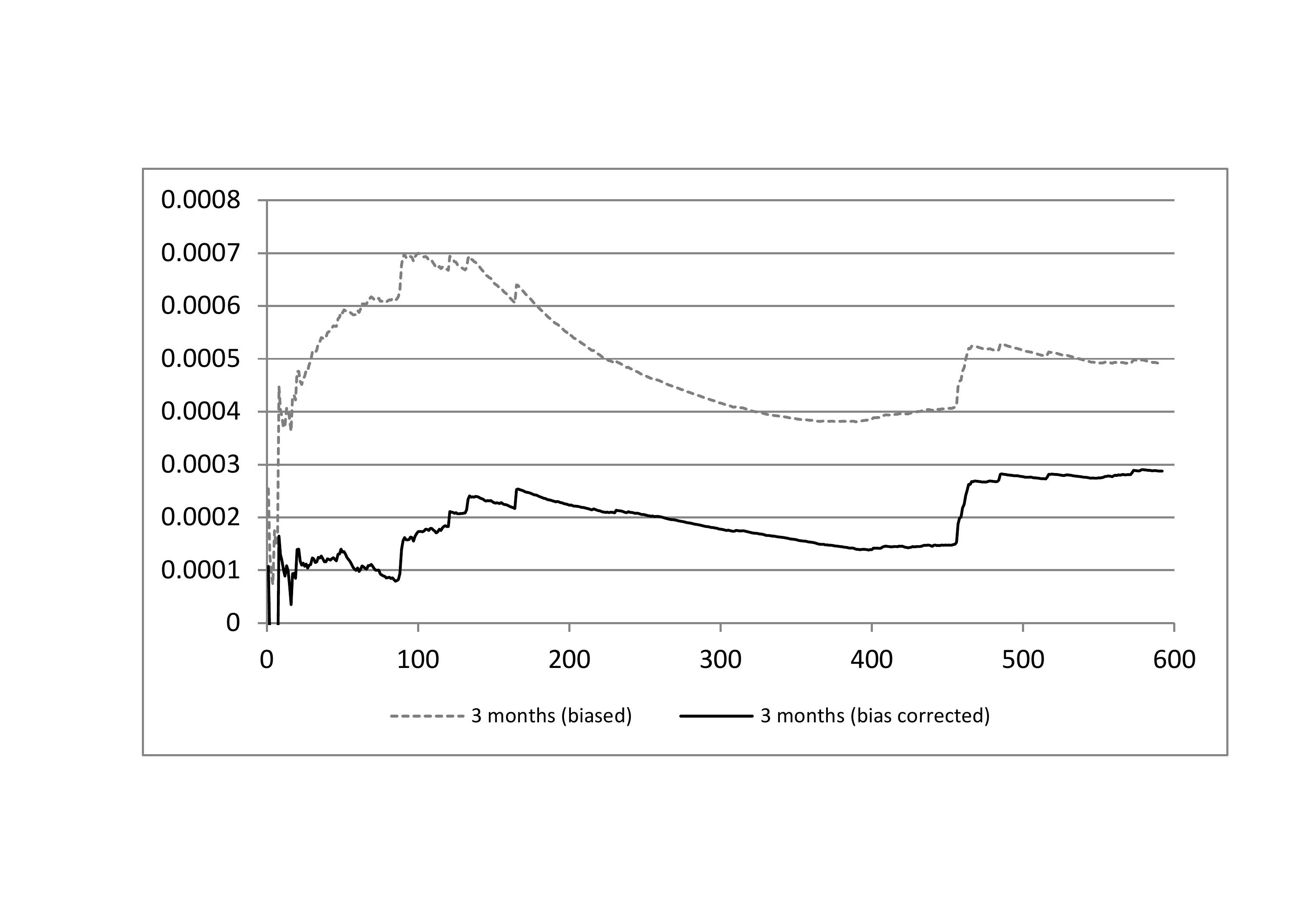}
\end{center}
\vspace{-2cm} 
\caption{Time series  $\widehat{s}_{ii}^{\rm bias}(K)$
and $\widehat{s}_{ii}(K)$, $K=1,\ldots, 600$,
 for maturity $m_i=3$ months and observations in
$ \{01/2000,\ldots, 05/2011\}$
on a weekly
grid $\Delta=1/52$.} \label{Figure 8}
\end{figure}

\begin{figure}[ht]
\vspace{-1cm}
\begin{center}
\includegraphics[width=\linewidth]{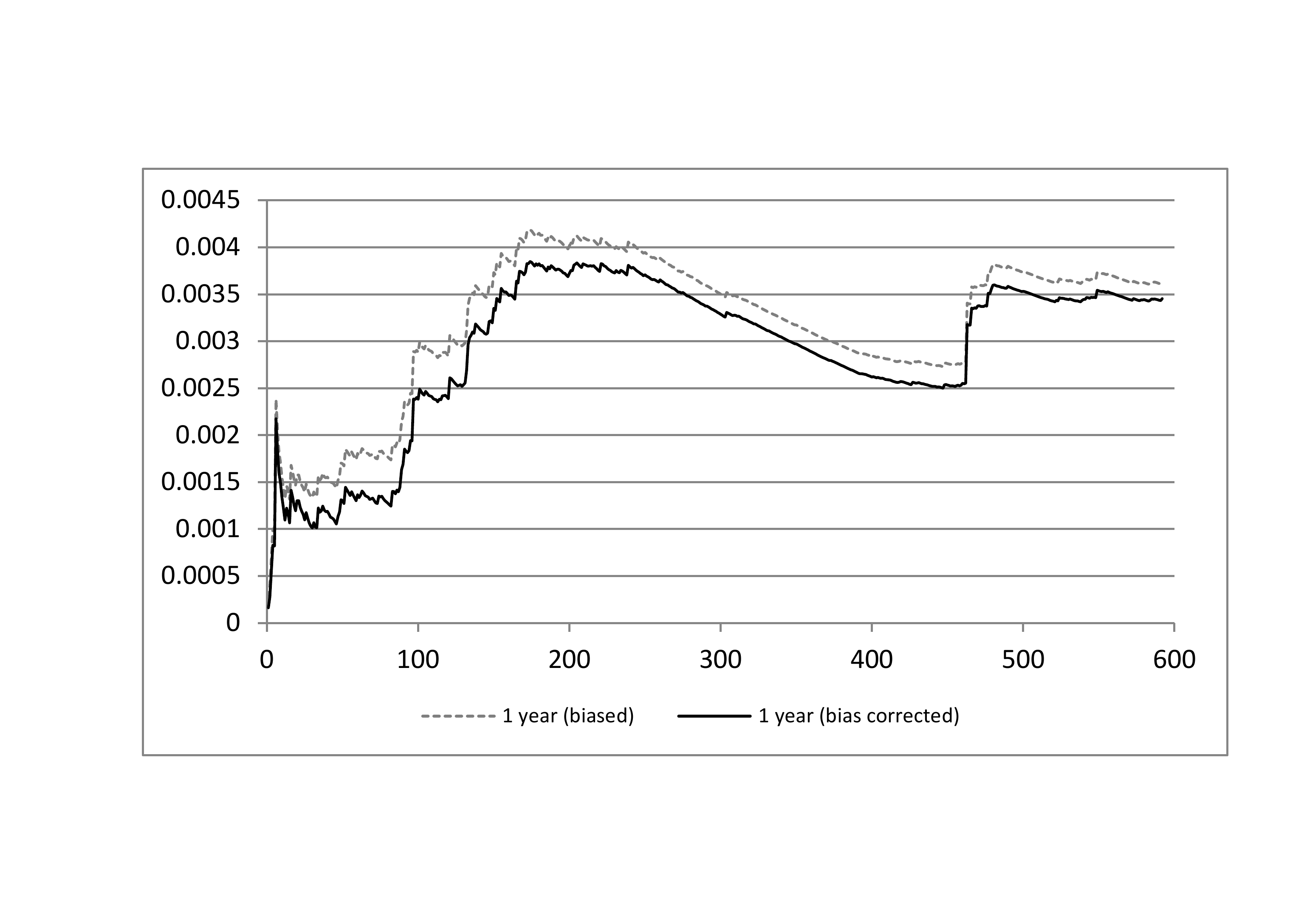}
\end{center}
\vspace{-2cm} 
\caption{Time series  $\widehat{s}_{ii}^{\rm bias}(K)$
and $\widehat{s}_{ii}(K)$, $K=1,\ldots, 600$,
 for maturity $m_i=1$ year and observations in
$ \{01/2000,\ldots, 05/2011\}$
on a weekly
grid $\Delta=1/52$.} \label{Figure 9}
%\end{figure}

%\begin{figure}[ht]
\vspace{-1cm}
\begin{center}
\includegraphics[width=\linewidth]{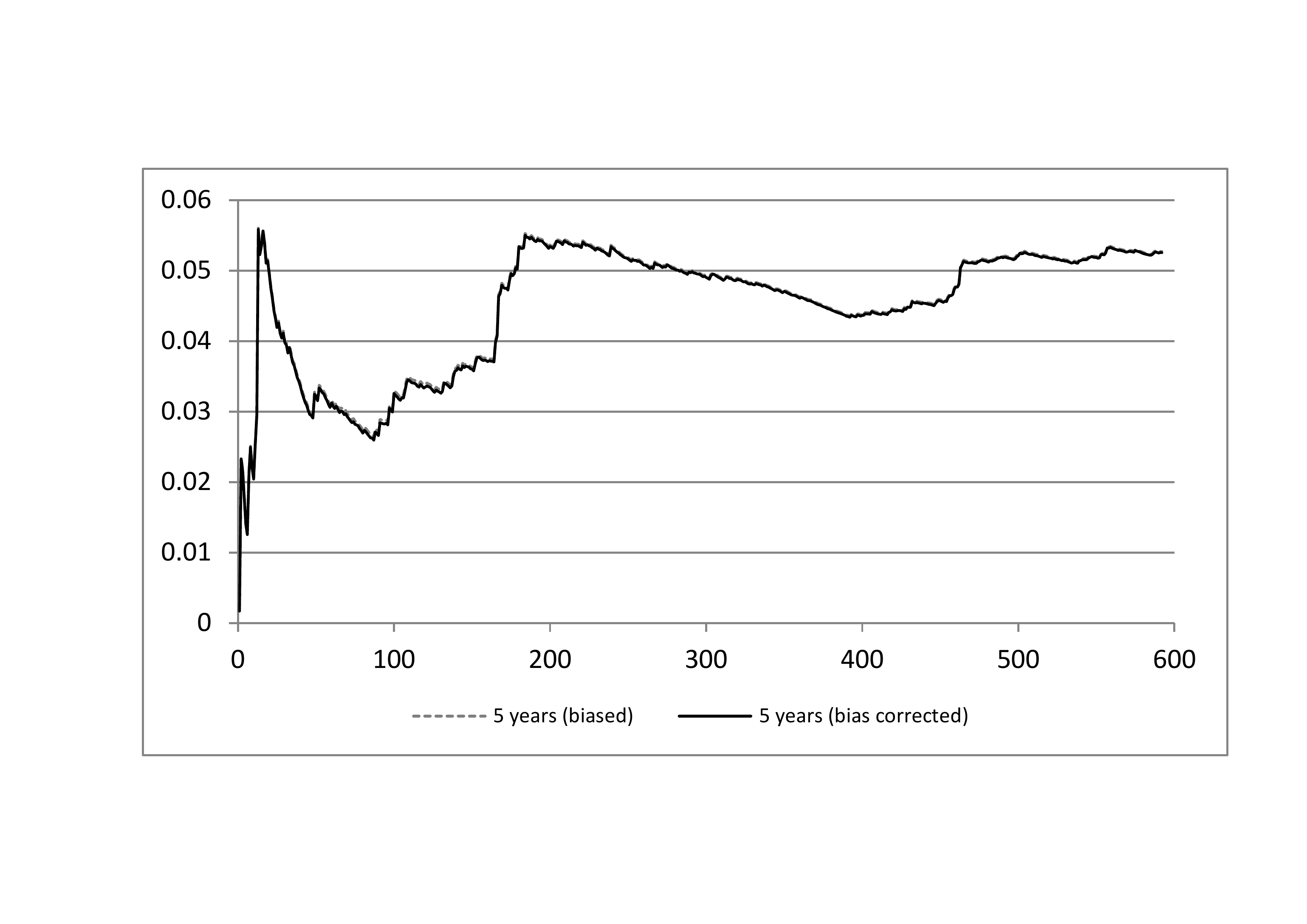}
\end{center}
\vspace{-2cm} 
\caption{Time series  $\widehat{s}_{ii}^{\rm bias}(K)$
and $\widehat{s}_{ii}(K)$, $K=1,\ldots, 600$,
 for maturity $m_i=5$ years and observations in
$ \{01/2000,\ldots, 05/2011\}$
on a weekly
grid $\Delta=1/52$.} \label{Figure 10}
\end{figure}

\begin{figure}[ht]
\vspace{-1cm}
\begin{center}
\includegraphics[width=\linewidth]{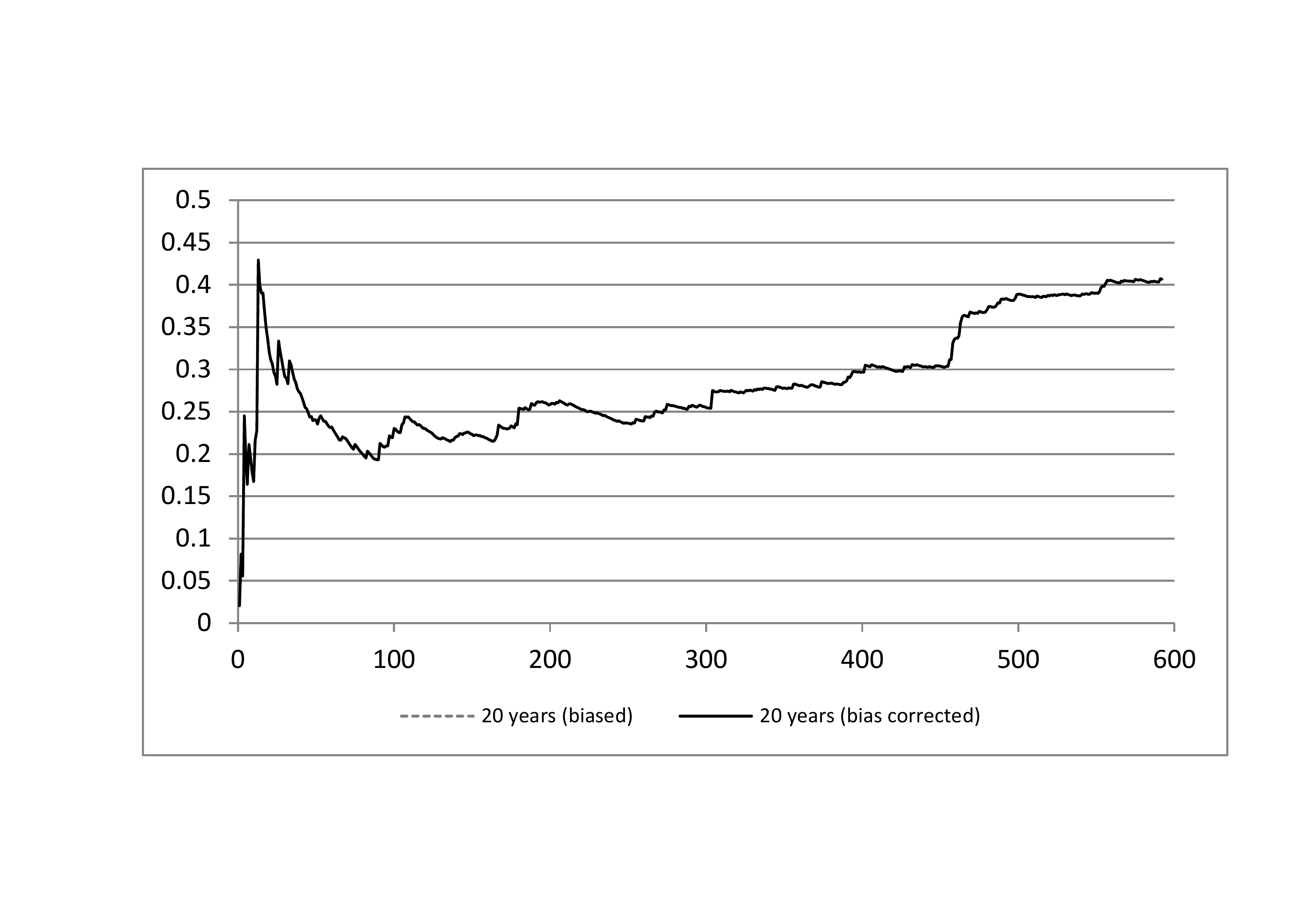}
\end{center}
\vspace{-2cm} 
\caption{Time series  $\widehat{s}_{ii}^{\rm bias}(K)$
and $\widehat{s}_{ii}(K)$, $K=1,\ldots, 600$,
 for maturity $m_i=20$ years and observations in
$ \{01/2000,\ldots, 05/2011\}$
on a weekly
grid $\Delta=1/52$.} \label{Figure 11}
%\end{figure}

%\begin{figure}[ht]
\vspace{-1cm}
\begin{center}
\includegraphics[width=\linewidth]{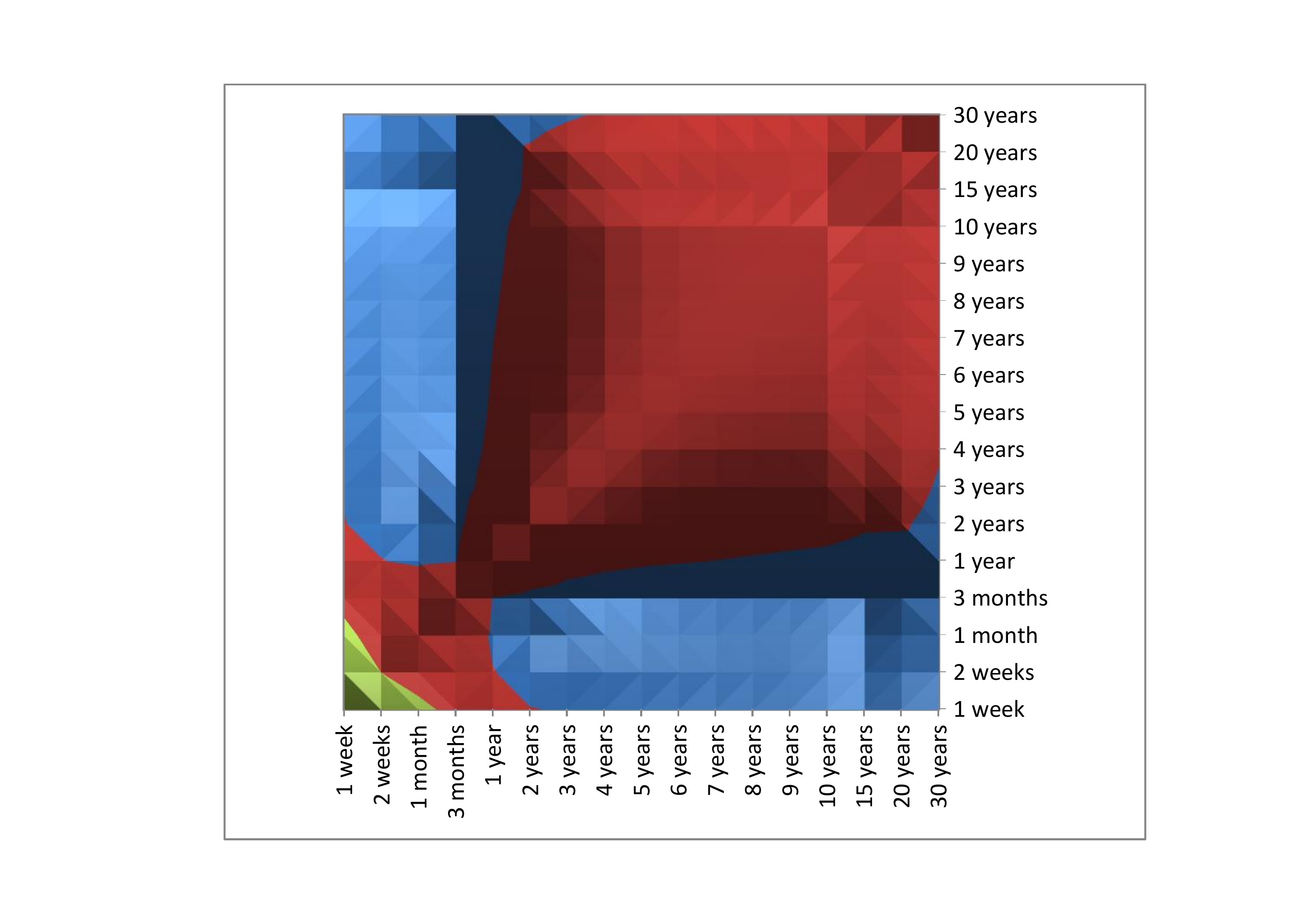}
\end{center}
\vspace{-1cm} 
\caption{Estimated matrix
$\widehat{\Xi}=(\widehat{\rho}_{ij})_{i,j=1,\ldots, d}$
from all observations in
$ \{01/2000,\ldots, 05/2011\}$
on a weekly
grid $\Delta=1/52$.} \label{Figure 12}
\end{figure}

Thus, for our analysis we choose the SAR for times to maturity
$m \in \{1/52, 1/26, 1/12, 1/4 \}$ and the Swiss government bond
for times to maturity $m \in \{1,2,3,4,5,6,7,8,9,10,15, 20, 30 \}$.
We choose time grid $\Delta=1/52$ (i.e.~a weekly time grid) and then
we calculate $\boldsymbol{\Upsilon}_t$ for our observations.
Note that we cannot directly calculate $\Upsilon_{t,m}=
m~Y(t,t+m)-(m+\Delta)~Y(t-\Delta,t+m)$ for all $m\in {\cal M}$ because
we have only a limited set of observed times to maturity. Therefore,
we make the following interpolation:
assume $m+\Delta \in (m, \widetilde{m}]$ for $m,\widetilde{m} \in {\cal M}$,
then approximate
\begin{equation*}
Y(t-\Delta,t+m)~\approx~
\frac{\widetilde{m}-(m+\Delta)}{\widetilde{m}-m} ~Y(t-\Delta, t+m-\Delta) + 
\frac{\Delta}{\widetilde{m}-m}~
Y(t-\Delta, t+\widetilde{m}-\Delta).
\end{equation*}
In Figure \ref{Figure 3} we give the time series of these 
estimated $(\boldsymbol{\Upsilon}_t)_t$ and in Figure \ref{Figure 4}
we give the component-wise ordered time series 
obtained from $(\boldsymbol{\Upsilon}_{t})_t$.
We observe that the volatility is increasing in the time to maturity due to
scaling with time to maturity. Using \eqref{choice sigma}
we calculate
\begin{equation*}
\sqrt{K} ~ C_{(K)} = \left(\left[
\varsigma(\mathbf{Y}_{\Delta k,-})^{-1}~
\boldsymbol{\Upsilon}_{\Delta k}\right]_j
\right)_{j=1,\ldots, d; ~k=1, \ldots, K}\in \R^{d\times K}
\end{equation*}
for our observations. In Figures \ref{Figure 5} and \ref{Figure 6}
we plot the time series $\Upsilon_{t,m}$ and 
$[\sqrt{K} ~ C_{(K)}]_m=
\Upsilon_{t,m}/h(Y(t-\Delta, t+m))$
for illustrative purposes only for maturities $m=1/52$ and $m=5$.
We observe that the scaling $\varsigma(\mathbf{Y}_{t,-})^{-1}$
gives more stationarity for short times to maturity, however in financial
stress periods it substantially increases the volatility of the observations,
see Figure \ref{Figure 5}.
For longer times to maturity one might discuss or even question the scaling
because it is less obvious whether it
is needed, see Figure \ref{Figure 6}. Next figures will show
that this scaling is also needed for longer times to maturity.
We then calculate the observed matrix
\begin{equation*}
\left(\widehat{s}_{ij}^{\rm bias}(K)\right)_{i,j=1,\ldots, d}=
\Delta^{-1}S_{(K)}(\mathbf{1})
\end{equation*}
as a function of the number of observations
$K$ (we set $\mathbf{1}=(1,\ldots, 1)'\in \R^d$). 
Moreover, we calculate the bias correction terms
given in \eqref{a_i calculation}-\eqref{c_i calculation}
where we simply replace the expected values on the right-hand
sides by the observations. Formulas 
\eqref{s_ii calculation}-\eqref{s_ij calculation} then provide
the estimates $\widehat{s}_{ij}(K)$ for $s_{ij}$ as a function
of the number of observations $K$. The bias correction term
is estimated by
\begin{equation*}
\widehat{\beta}_{ij}(K)=
\widehat{s}_{ij}^{\rm bias}(K)
-\widehat{s}_{ij}(K).
\end{equation*}
We expect that for short times to maturity the bias correction term is larger due to more
dramatic drifts. The results for selected times to maturity 
$m\in \{1/52, 1/4, 1, 5, 20\}$
are presented in Figures \ref{Figure 7}-\ref{Figure 11}. Let us 
comment these figures:
\begin{itemize}
\item Times to maturity in the set ${\cal M}_1=\{1/52,1/26, 1/12\}$ 
look similar to $m=1/52$ (Figure \ref{Figure 7});
${\cal M}_2=\{1/4\}$ corresponds to Figure \ref{Figure 8};
times to maturity in the set ${\cal M}_3=
\{1,2,3,4,5,6,7,8,9,10,15\} $ look similar to $m=1,5$
(Figures \ref{Figure 9}-\ref{Figure 10});
times to maturity $m\in {\cal M}_4=\{20,30\}$ look similar to $m=20$
(Figure \ref{Figure 11}).
\item Times to maturity in ${\cal M}_1 \cup {\cal M}_3$ seem to have converged,
for ${\cal M}_2$ the convergence picture is distorted by the
last financial crisis, where volatilities relative to yields have
substantially increased, see also Figure \ref{Figure 5}. One might
ask whether during financial crisis we should apply a different
scaling (similar to regime switching models). For ${\cal M}_4$ the
convergence picture suggest that we should probably study longer time
series (or scaling should be done differently).
Concluding, this supports the choice of the function $h$
in \eqref{h function}. Only long times to maturity 
$m \in {\cal M}_4$ might suggest a different scaling.
\item For times to maturities in ${\cal M}_3 \cup {\cal M}_4$ we observe that
the bias term given in \eqref{asymptotic result} is negligible, 
see Figures \ref{Figure 9}-\ref{Figure 11},
that is, $\Delta
=1/52$ is sufficiently small for times to maturity $m\ge 1$. For
times to maturities in ${\cal M}_1 \cup {\cal M}_2$ it is however
essential that we do a bias correction, see Figure 
\ref{Figure 7}-\ref{Figure 8}. This comes from the fact that for
small times to maturity the bias term is driven by 
$\mathbf{z}$ in $f_\Lambda(\mathbf{z},\mathbf{y})$ which then is
of similar order as $s_{ii}$.
\end{itemize}
In Table \ref{Table 1} we present the resulting
estimated matrix $\widehat{\Sigma}_\Lambda(\mathbf{1})
=(\widehat{s}_{ij}(K))_{i,j=1,\ldots, d}$ which is based on all 
observations in $ \{01/2000,\ldots, 05/2011\}$. We observe that
the diagonal $\widehat{s}_{ii}(K)$ is an increasing function
in the time to maturity $m_i$. 
Therefore, in order to further analyze this
matrix, we normalize it as follows (as a correlation matrix)
\begin{equation*}
\widehat{\Xi}=(\widehat{\rho}_{ij})_{i,j=1,\ldots, d}
=\left(\frac{\widehat{s}_{ij}(K)}{\sqrt{\widehat{s}_{ii}(K)}
\sqrt{\widehat{s}_{jj}(K)}}\right)_{i,j=1,\ldots, d}.
\end{equation*}
Now all the entries $\widehat{\rho}_{ij}$ live on the same
scale and the result is presented in Figure \ref{Figure 12}.
We observe two different structures, one for times to maturity
less than 1 year, i.e.~$m\in \widetilde{\cal M}_1
={\cal M}_1 \cup {\cal M}_2$, 
and one for times to maturity $m\in 
\widetilde{\cal M}_2=
{\cal M}_3 \cup {\cal M}_4$.
The former times to maturity $m\in \widetilde{\cal M}_1$
were modeled using the observations
from the SAR, the latter $m\in \widetilde{\cal M}_2$
with observations from the Swiss
government bond. This separation shows that these two data
sets are not completely compatible which gives some
,,additional independence'' (diversification) between 
$\widetilde{\cal M}_1$ and $\widetilde{\cal M}_2$.
If we calculate the eigenvalues of $\widehat{\Xi}$
we observe that the first 5 eigenvalues explain 95\% of the
total observed cross-sectional volatility (we have a $d=17$ dimensional
space). Thus, a principal component analysis says that
we should at least choose a 5-factor model. These are more factors
than typically stated in the literature (see Brigo-Mercurio \cite{BM},
Section 4.1). The reason therefore 
is again that the short end $\widetilde{\cal M}_1$
and the long end $\widetilde{\cal M}_2$ of the estimated yield curve
behave more independently due to different choices of the data
(see also Figure \ref{Figure 12}). If we restrict this principal
component analysis to $\widetilde{\cal M}_2$ we find the classical
result that a 3-factor model explains 95\% of the observed
cross-sectional volatility.

\begin{figure}[ht]
\vspace{-1cm}
\begin{center}
\includegraphics[width=\linewidth]{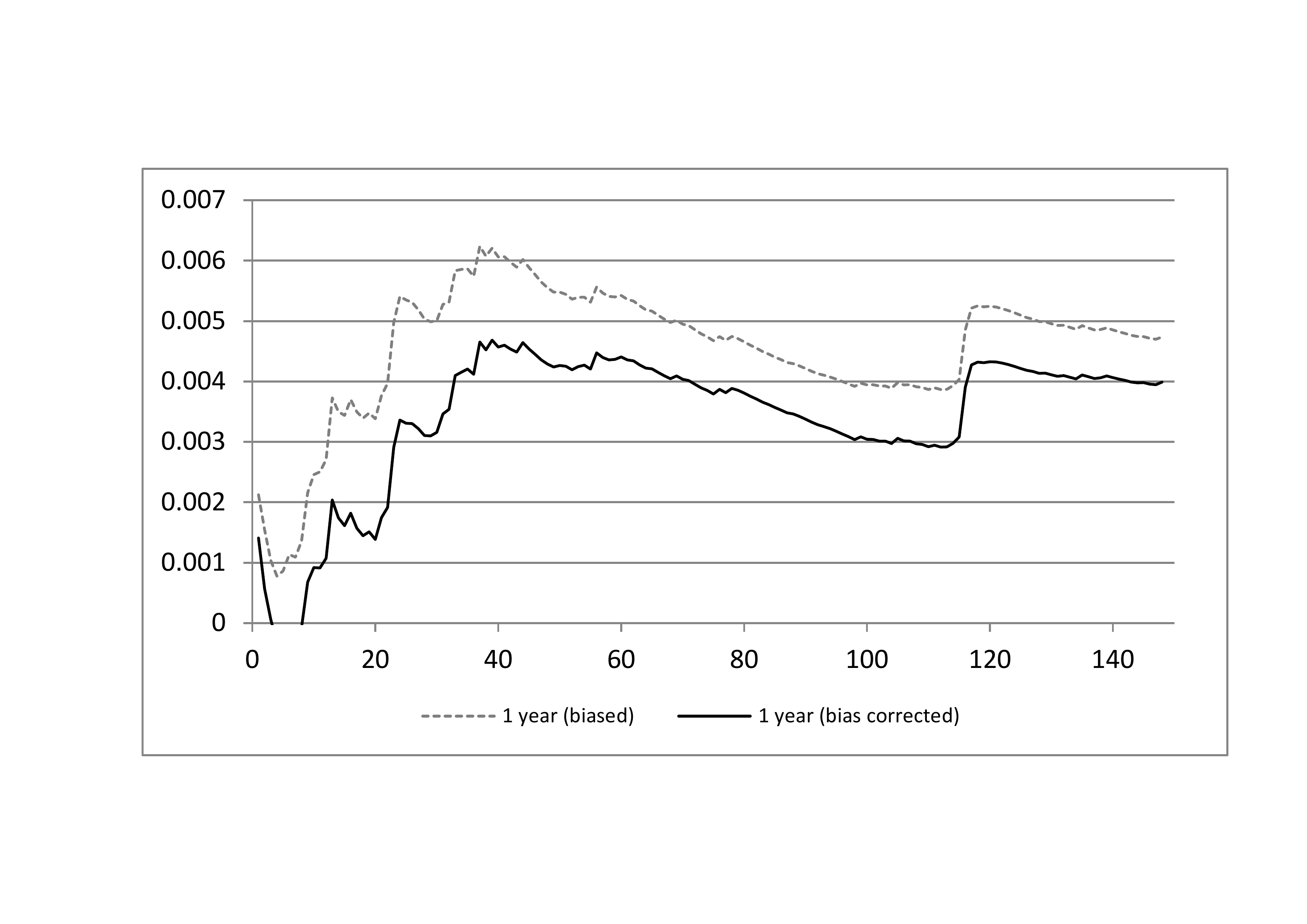}
\end{center}
\vspace{-2cm} 
\caption{Time series  $\widehat{s}_{ii}^{\rm bias}(K)$
and $\widehat{s}_{ii}(K)$, $K=1,\ldots, 600$,
 for maturity $m_i=1$ year and observations in
$ \{01/2000,\ldots, 05/2011\}$
on a monthly
grid $\Delta=1/12$.} \label{Figure 9_3}
\end{figure}

In the next step we analyze the assumption
of the independence of $\Sigma_\Lambda(\mathbf{1})=\Lambda \Lambda'
=(s_{ij})_{i,j=1,\ldots, d}$ from the grid size $\Delta$. Similar
to the analysis above we estimate $\Sigma_\Lambda(\mathbf{1})$ for the
grid sizes $\Delta=1/52, 1/26, 1/13, 1/4$ (weekly, bi-weekly,
4-weekly, quarterly grid size). The first observation is that the
bias increases with increasing $\Delta$ (for illustrative purposes
one should compare Figure \ref{Figure 9} with $m=1$ and $\Delta=1/52$
and Figure \ref{Figure 9_3} with $m=1$ and $\Delta=1/4$). Of course,
this is exactly the result expected.

In Table \ref{Table 2} we give the differences between
the estimated matrices $\widehat{\Sigma}_\Lambda(\mathbf{1})
=(\widehat{s}_{ij}(K))_{i,j=1,\ldots, d}$ on the 
weekly grid $\Delta=1/52$ versus the
estimates on a quarterly grid $\Delta =1/4$ (relative to the 
estimated values on the quarterly grid). Of course, we can only
display these differences for times to maturity $m\in {\cal M}_2
\cup \widetilde{\cal M}_2$ because in the latter model the times
to maturity in ${\cal M}_1$ do not exist. We observe rather small
differences within $\widetilde{\cal M}_2$ which supports the
independence assumption from the choice of $\Delta$ within
the Swiss government bond yields. For the SAR in ${\cal M}_2$ this
picture does not entirely hold true which has also to do with
the fact that the model does not completely fit to the data,
see Figure \ref{Figure 8}. Thus, we only observe larger difference
for covariances that have a bigger difference in times to maturity
compared. The pictures for $\Delta=1/26, 1/13$
are quite similar which justify our independence choice.

\begin{conclusions}~\label{conclusions}\end{conclusions}
We conclude that the independence assumption
of $\Sigma_\Lambda(\mathbf{1})$ from $\Delta$ 
is not violated by our observations
and that the bias terms $\widehat{\beta}_{i,j}(K)$ are negligible
for maturities $m_i,m_j \in \widetilde{\cal M}_2$ and time grids
$\Delta=1/52, 1/16, 1/13$, therefore we can directly work with
model \eqref{model kappa} to predict future yields for
times to maturity in $\widetilde{\cal M}_2$.

\subsection{Back-testing and market price of risk}
\label{subsection back-testing}
In this subsection we back-test our model against the
observations. We therefore choose a fixed-term annuity with nominal
payments of size 1 at maturity dates $m\in {\cal M}_3$. The present
value of this annuity at time $t$ is given by
\begin{equation*}
\pi_t = \sum_{m \in {\cal M}_3} P(t,t+m)=
\sum_{m \in {\cal M}_3} \exp \left\{- m~ Y(t,t+m) \right\}
\approx \sum_{m \in {\cal M}_3} 1- m ~Y(t,t+m)~ \stackrel{\rm def.}{=}~
\widetilde{\pi}_t.
\end{equation*}
Our back-testing setup is such that we try to predict  $\widetilde{\pi}_{t}$ based on the observations ${\cal F}_{t-\Delta }$ and then (one period later) we compare this forecast with the realization
of $\widetilde{\pi}_{t}$.  In view of Conclusions \ref{conclusions} we
directly work with $C_{(K)}$ for small time grids $\Delta$
(for $t=\Delta(K+1)$). Moreover, the Taylor approximation $\widetilde{\pi}_t$ to 
${\pi}_t$ is used in order to avoid (time-consuming) simulations. Here a first order Taylor expansion is sufficient since the portfolio's variance will be -- due to high positive correlation -- quite large in comparison to possible second order -- drift like -- correction terms. Such an approximation does not work for short-long portfolios.

\begin{figure}[ht]
\vspace{-1cm}
\begin{center}
\includegraphics[width=\linewidth]{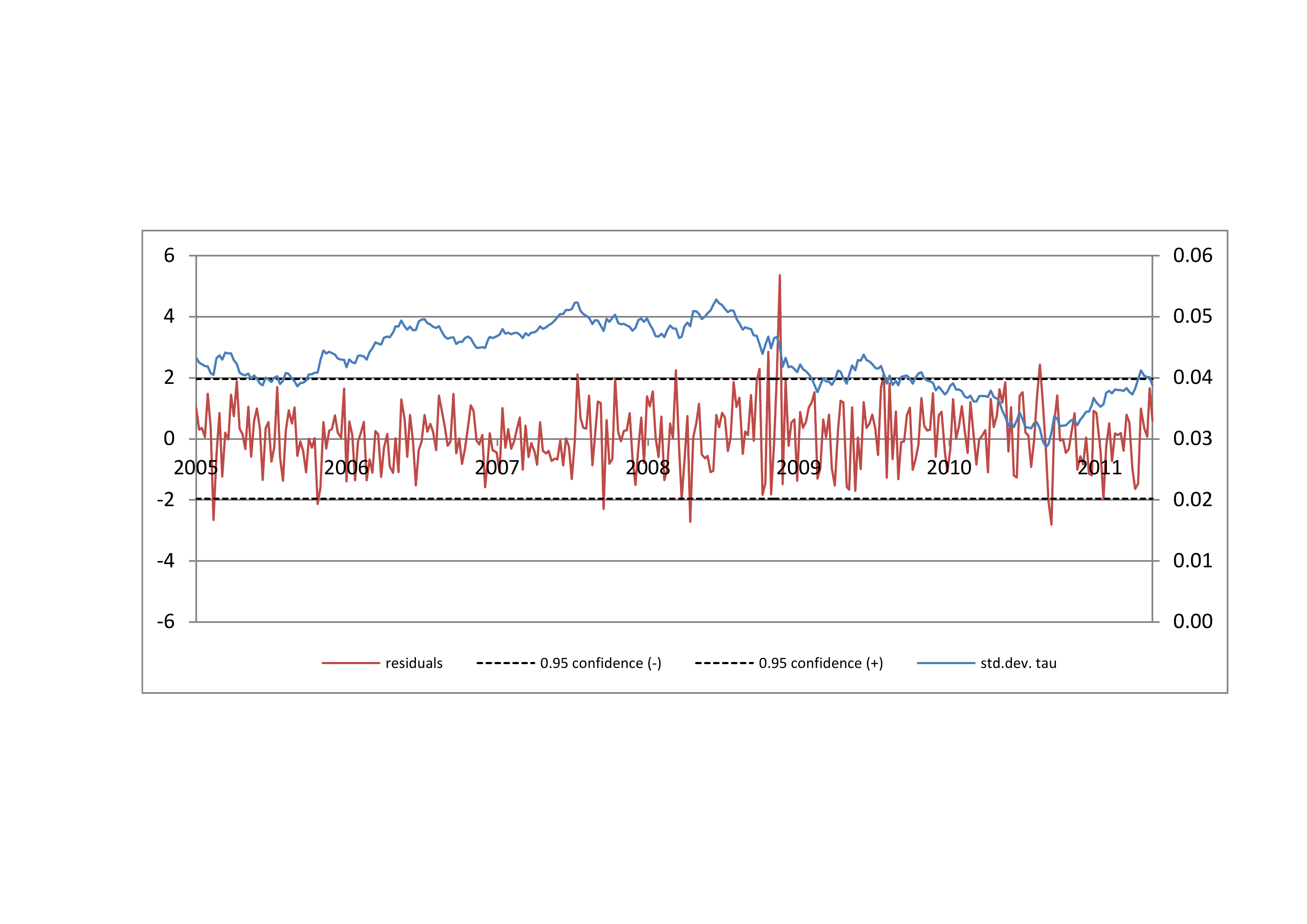}
\end{center}
\vspace{-2cm} 
\caption{Time series  of residuals $z^\ast_t$
for $t \in \{01/2005,\ldots, 05/2011\}$
on a weekly grid $\Delta=1/52$.
The axis on the right-hand side displays the time series of $\tau_{t-\Delta}$.
} \label{residuals 1}
%\end{figure}

%\begin{figure}[ht]
\vspace{-1cm}
\begin{center}
\includegraphics[width=\linewidth]{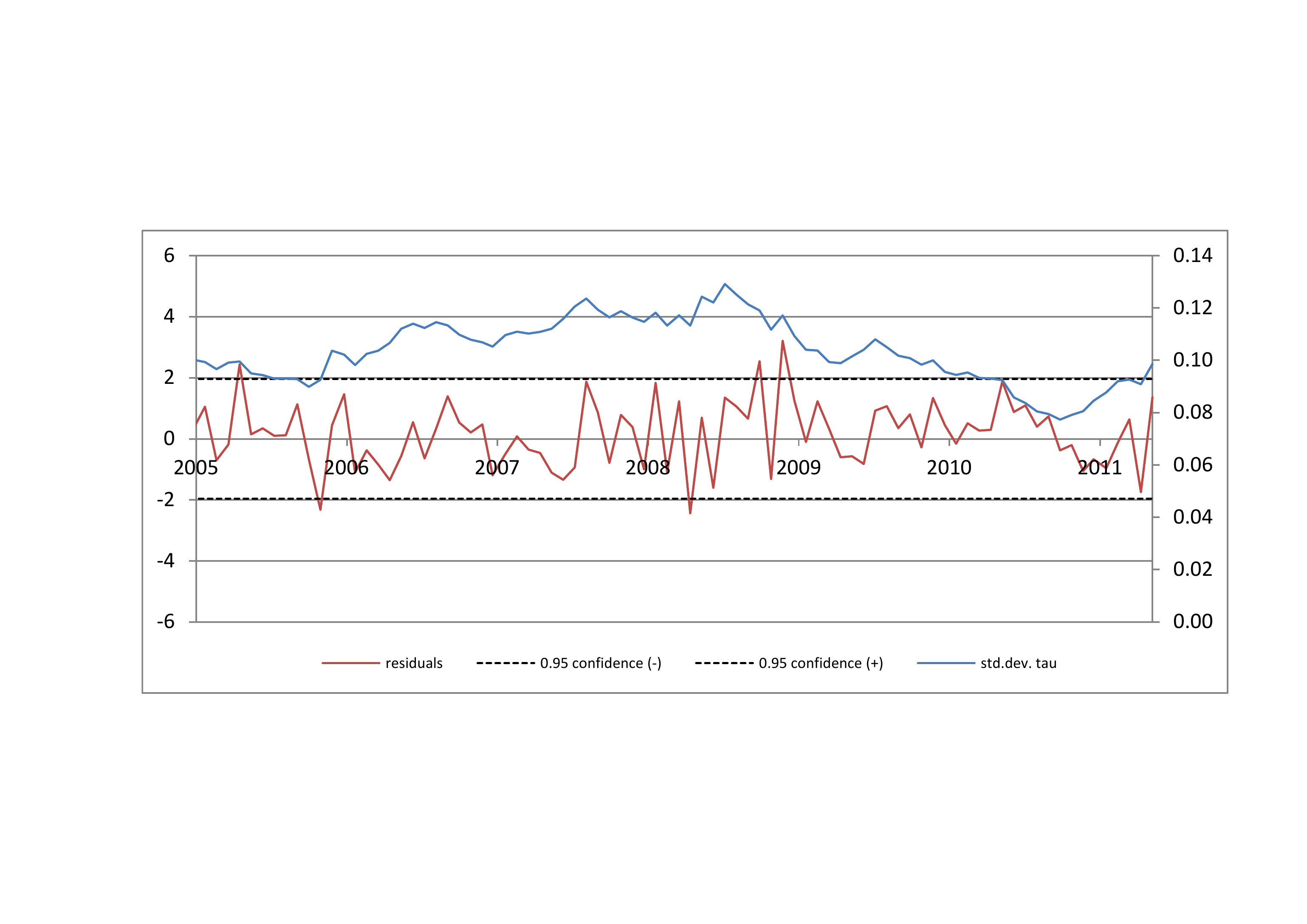}
\end{center}
\vspace{-2cm} 
\caption{Time series  of residuals $z^\ast_t$
for $t \in \{01/2005,\ldots, 05/2011\}$
on a monthly grid $\Delta=1/12$.
The axis on the right-hand side displays the time series of $\tau_{t-\Delta}$.
} \label{residuals 2}
\end{figure}

\begin{figure}[ht]
\vspace{-1cm}
\begin{center}
\includegraphics[width=\linewidth]{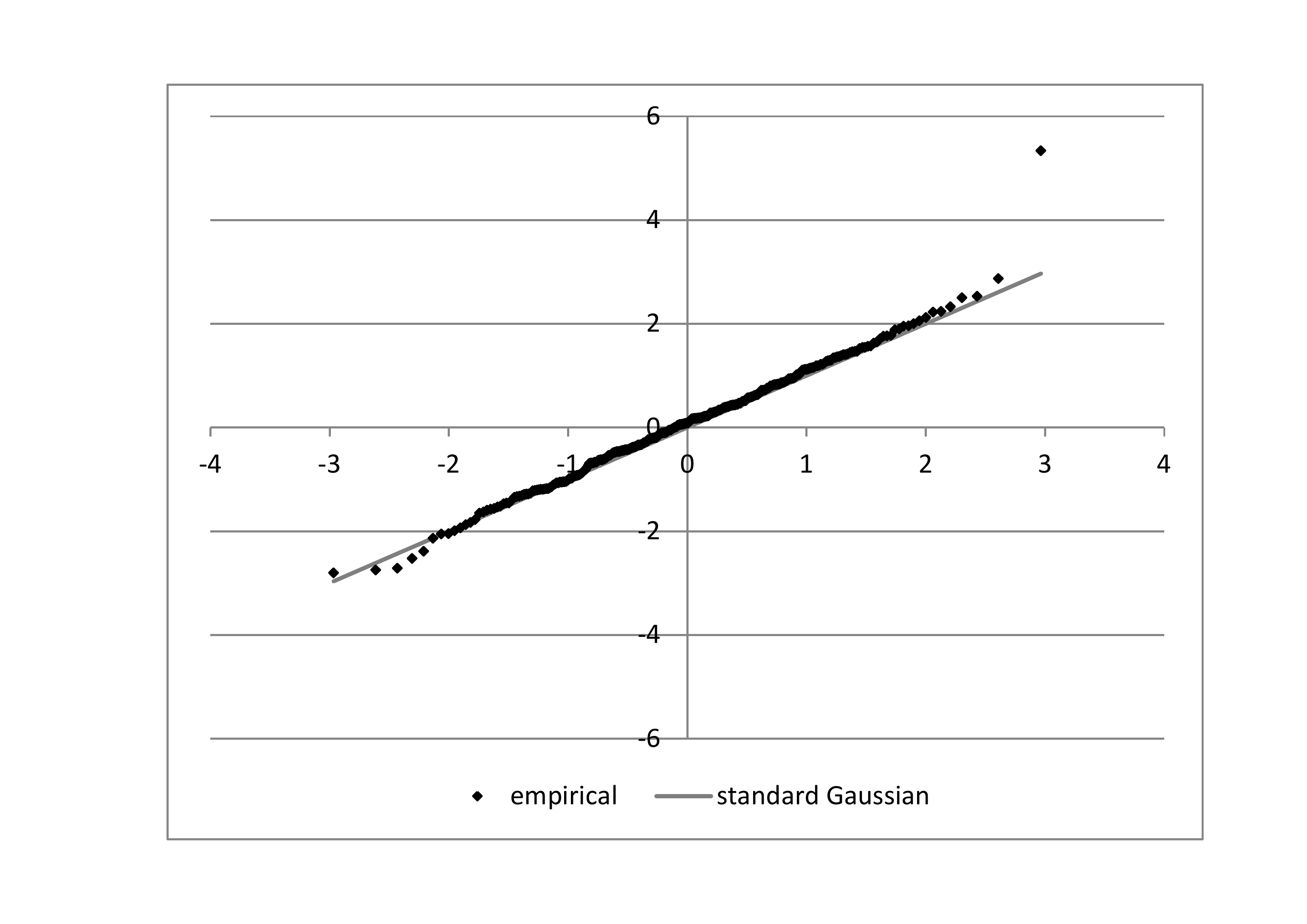}
\end{center}
\vspace{-1cm} 
\caption{Q-Q-plot of the residuals $(z^\ast_t)_t$ against the
standard Gaussian distribution for $\Delta=1/52$.
} \label{QQ Plot}
%\end{figure}

%\begin{figure}[ht]
\vspace{-1cm}
\begin{center}
\includegraphics[width=\linewidth]{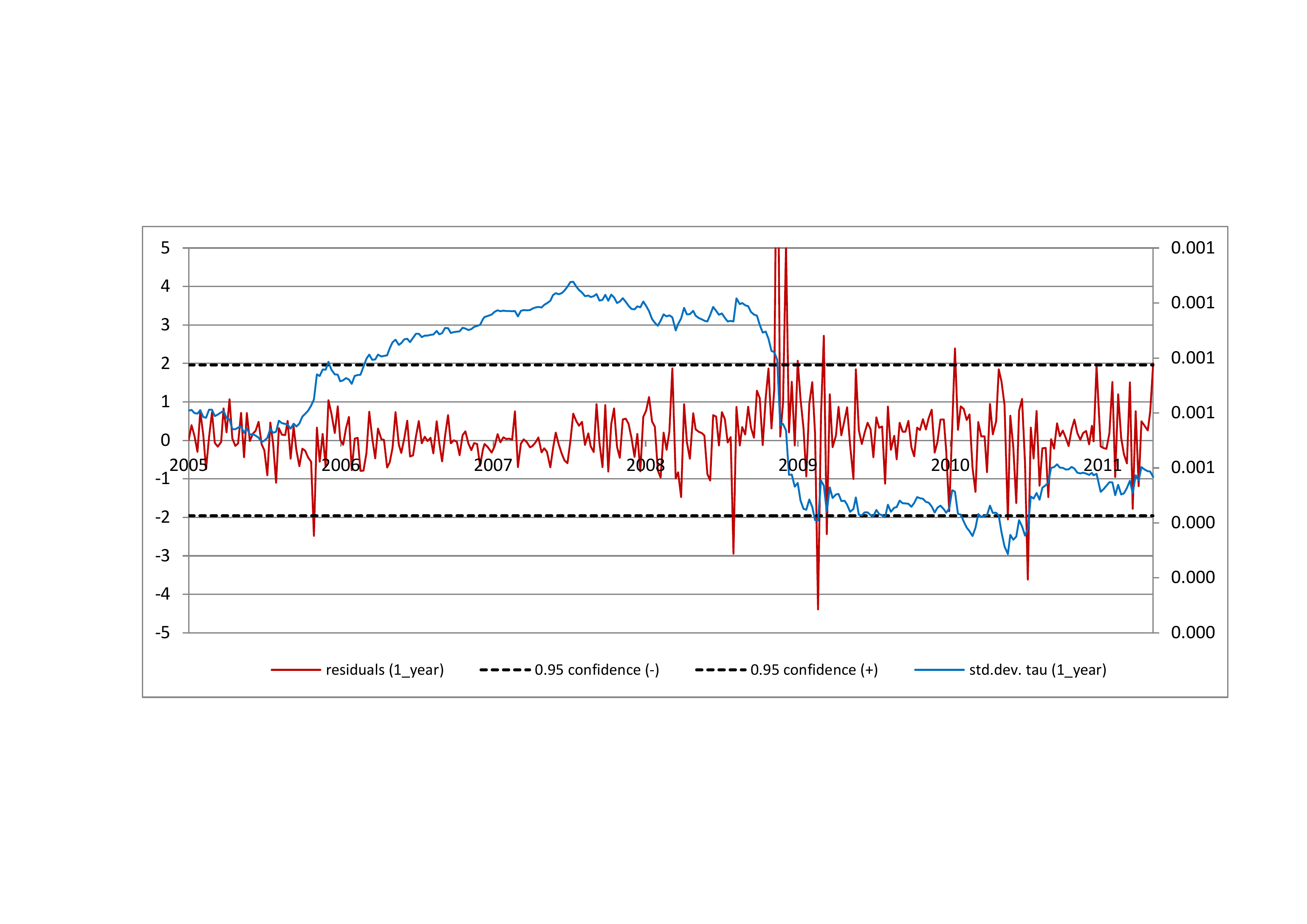}
\end{center}
\vspace{-2cm} 
\caption{Time series  of residuals $z^\ast_{m,t}$
for time to maturity $m=1$ and
$t \in \{01/2005,\ldots, 05/2011\}$
on a weekly grid $\Delta=1/52$.
The axis on the right-hand side displays the time series of 
$\tau_{m,t-\Delta}$.
} \label{maturity 1}
\end{figure}

For the approximation (under $\p^\ast$)
\begin{equation*}
\boldsymbol{\Upsilon}_{t}|_{{\cal F}_{t-\Delta}}
~\stackrel{(d)}{\approx}~
\boldsymbol{\kappa}_{t}(\mathbf{Y}(t-\Delta, t),
\mathbf{Y}_{t,-})|_{{\cal F}_{t-\Delta}},
\end{equation*}
we obtain an approximate forecast to $\widetilde{\pi}_t$ given by
(denote the cardinality of ${\cal M}_3$ by $d_3$)
\begin{eqnarray}\nonumber
\widetilde{\widetilde{\pi}}_t|_{{\cal F}_{t-\Delta}}
&=&d_3-
\sum_{m \in {\cal M}_3} (m+\Delta)~Y(t-\Delta,t+m)
+d_3 ~\Delta Y(t-\Delta, t)\\\label{life portfolio}
&&\qquad 
-\frac{1}{2}~\mathbf{1}_{{\cal M}_3}'~{\rm sp}\left(S_{(K)}
(\mathbf{Y}_{t,-})\right)
-\mathbf{1}_{{\cal M}_3}'~\varsigma(\mathbf{Y}_{t,-})C_{(k)}\mathbf{W}^\ast_t~
\big|_{{\cal F}_{t-\Delta}},
\end{eqnarray}
where $\mathbf{1}_{{\cal M}_3}=(1_{\{1\in {\cal M}_3\}},\ldots, 
1_{\{d\in {\cal M}_3\}})'\in \R^d$.
Thus, the conditional distribution of $\widetilde{\widetilde{\pi}}_t$
under $\p^\ast$, given ${\cal F}_{t-\Delta}$, is a Gaussian distribution
with conditional mean and conditional variance given by
\begin{eqnarray*}
\mu^\ast_{t-\Delta}&=&
d_3-
\sum_{m \in {\cal M}_3} (m+\Delta)~Y(t-\Delta,t+m)
+d_3 ~\Delta Y(t-\Delta, t)
-\frac{1}{2}~\mathbf{1}_{{\cal M}_3}'~{\rm sp}\left(S_{(K)}
(\mathbf{Y}_{t,-})\right),\\
\tau_{t-\Delta}^2&=&
\mathbf{1}_{{\cal M}_3}'~
S_{(K)}
(\mathbf{Y}_{t,-})~\mathbf{1}_{{\cal M}_3}.
\end{eqnarray*}
We calculate these conditional moments for
$t \in \{01/2005, \ldots, 05/2011 \}$ based on the $\sigma$-fields
${\cal F}_{t-\Delta}$
generated by the data in 
$\{01/2000, \ldots, t-\Delta \}$, for $\Delta=1/52, 1/12$ (weekly and
monthly grid). From these we can calculate the observable residuals
\begin{equation*}
z^\ast_t = \frac{\widetilde{\pi}_t-\mu^\ast_{t-\Delta}}{\tau_{t-\Delta}}.
\end{equation*}
The sequence of these observable residuals should 
approximately look like an i.i.d.~standard Gaussian distributed
sequence. The result for $\Delta=1/52$ is given in Figure \ref{residuals 1}
and for $\Delta=1/12$ in Figure \ref{residuals 2}.
At the first sight this sequence $(z^\ast_t)_t$ seems to fulfill
these requirements, thus the out-of-sample back-testing provides the
required results. In Figure \ref{QQ Plot} we also provide
the Q-Q-plot for the residuals $(z^\ast_t)_t$
against the standard Gaussian distribution for
$\Delta =1/52$. Also in this plot we observe a good fit, except
for the tails of the distribution. This suggests that one may
relax the Gaussian assumption on $\boldsymbol{\varepsilon}_t^\ast$
by a more heavy-tailed model (this can also be seen
in Figure \ref{residuals 1} where we a few outliers). We have already
mentioned this in Section \ref{modelling_na} but for this exposition we keep
the Gaussian assumption.

If we calculate the auto-correlation for time lag $\Delta$ 
between the residuals $z^\ast_t$
 we obtain 5\% which is a convincingly small value.
This supports the assumption having independent residuals. The
same holds true if we consider the 
auto-correlation for time lag $\Delta$ between the
absolute values $|z^\ast_t|$ of the residuals resulting in 11\%.
The only observation which may contradict the i.i.d.~assumption
is that we observe slight  clustering in Figure \ref{residuals 1}.
This non-stationarity might have to do with that we calculate the
residuals under the equivalent martingale measure $\p^\ast$,
however we make the observations under the real world probability
measure $\p$. If these measures coincide the statements are the same.

The classical approach is that one assumes that the two probability
measures are equivalent, i.e.~$\p^\ast \sim \p$, with density process
\begin{equation}\label{market-price of risk}
\xi_{t} = \prod_{s=1}^{t/\Delta} \exp \left\{ -\frac{1}{2}\left\|
\boldsymbol{\lambda}_{\Delta s} \right\|^2+
\boldsymbol{\lambda}_{\Delta s}
~ \boldsymbol{\varepsilon}_{\Delta s} \right\},
\end{equation}
with $\boldsymbol{\varepsilon}_t$ is independent
of ${\cal F}_{t-\Delta}$, ${\cal F}_{t}$-measurable 
and a $t/\Delta$-dimensional
standard Gaussian random vector
with independent components under $\p$. 
Moreover, it is assumed that $\boldsymbol{\lambda}_{t}$ is 
$d$-dimensional and previsible, i.e.~${\cal F}_{t-\Delta}$-measurable.
Note that this density process
$(\xi_t)_t$ is a strictly positive and normalized $(\p,\F)$-martingale.
For any $\p^\ast$-integrable and ${\cal F}_t$-measurable random variable
$X_t$ we have, $\p$-a.s.,
\begin{equation*}
\E^\ast_{t-\Delta} \left[X_t\right]=
\frac{1}{\xi_{t-\Delta}}~\E_{t-\Delta} \left[\xi_t X_t\right].
\end{equation*}
This implies that
\begin{equation*}
\boldsymbol{\varepsilon}_{t}-\boldsymbol{\lambda}_{t}
~\stackrel{(d)}{=}~\boldsymbol{\varepsilon}^\ast_t
\quad \text{under $\p^\ast_{t-\Delta}$.}
\end{equation*}
$\boldsymbol{\lambda}_{t}$ is called market price
of risk at time $t$ and reflects
the difference between $\p^\ast_{t-\Delta}$
and $\p_{t-\Delta}$. Under Model Assumptions \ref{Model Assumptions 1}
we then obtain under the real world probability measure $\p$
\begin{equation*}
\boldsymbol{\Upsilon}_t
= \Delta\left[-\mathbf{Y}(t-\Delta,t)+\frac{1}{2}
~{\rm sp}
(\Sigma_{{\Lambda}} 
({\mathbf{Y}_{t, -}}))
\right]+ \sqrt{\Delta}~
\varsigma({\mathbf{Y}_{t, -}})~{\Lambda} 
~ \boldsymbol{\lambda}_{t}
+ \sqrt{\Delta}~
\varsigma({\mathbf{Y}_{t, -}})~{\Lambda} 
~ \boldsymbol{\varepsilon}_{t},
\end{equation*}
i.e.~we have a change of drift given by $
\sqrt{\Delta}~
\varsigma({\mathbf{Y}_{t, -}})~{\Lambda} 
~ \boldsymbol{\lambda}_{t}$.
Thus, under the (conditional) real world probability
measure $\p_{t-\Delta}$ the approximate forecast
$\widetilde{\widetilde{\pi}}_t$ has a Gaussian distribution
with conditional mean and conditional covariance given by
\begin{equation*}
\mu_{t-\Delta}=
\mu^\ast_{t-\Delta}
-\sqrt{\Delta}~\mathbf{1}_{{\cal M}_3}'~\varsigma
(\mathbf{Y}_{t,-})~\Lambda~\boldsymbol{\lambda}_t\qquad
\text{ and } \qquad
\tau_{t-\Delta}^2=
\mathbf{1}_{{\cal M}_3}'~
S_{(K)}
(\mathbf{Y}_{t,-})~\mathbf{1}_{{\cal M}_3}.
\end{equation*}
For an appropriate choice of the market price of risk 
$\boldsymbol{\lambda}_{t}$ we obtain residuals
\begin{equation*}
z_t = \frac{\widetilde{\pi}_t-\mu_{t-\Delta}}{\tau_{t-\Delta}},
\end{equation*}
which should then form an i.i.d.~standard Gaussian distributed
sequence under the real world probability measure $\p$.

\begin{figure}
\vspace{-1cm}
\begin{center}
\includegraphics[width=\linewidth]{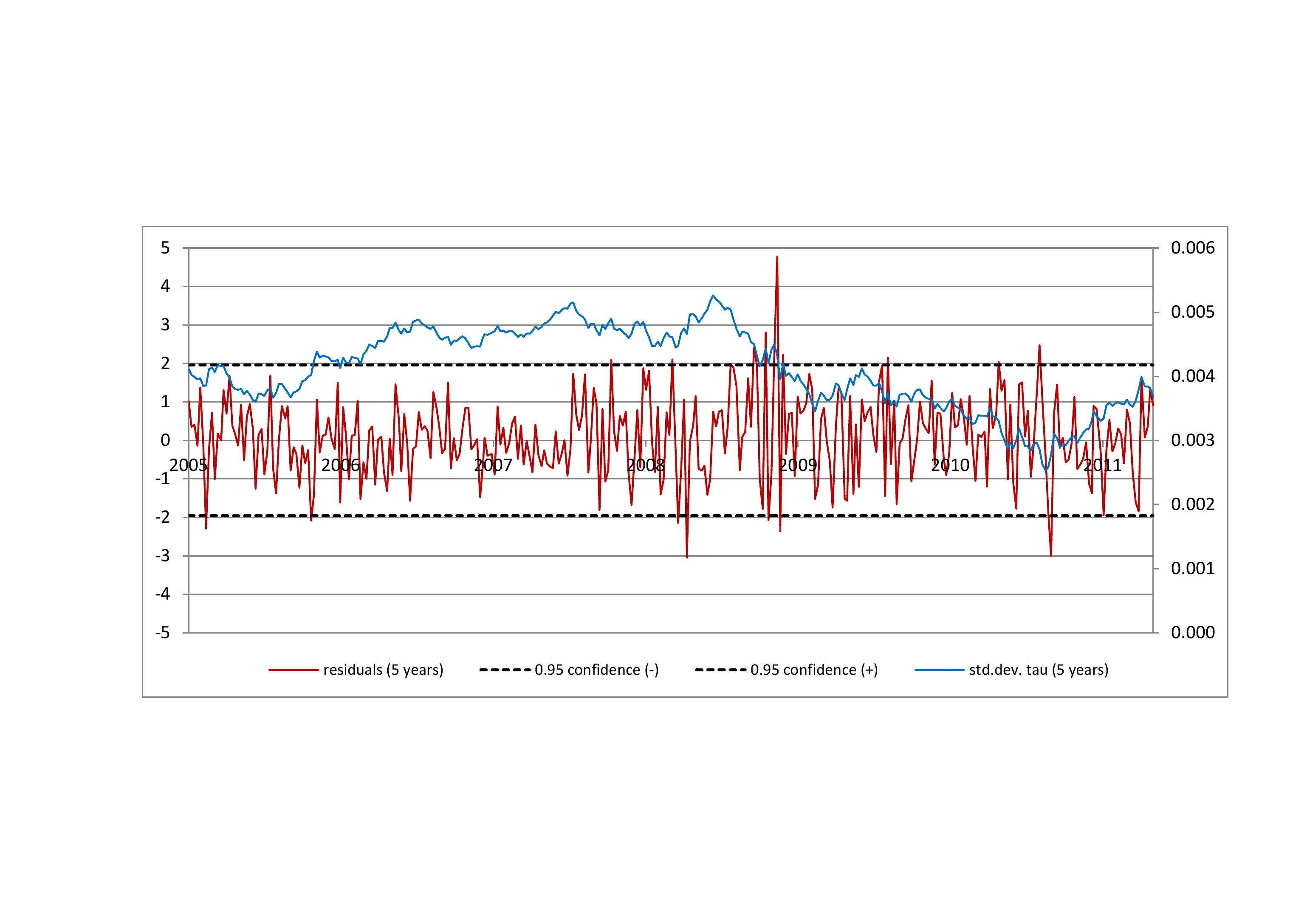}
\end{center}
\vspace{-2cm} 
\caption{Time series  of residuals $z^\ast_{m,t}$
for time to maturity $m=5$ and
$t \in \{01/2005,\ldots, 05/2011\}$
on a weekly grid $\Delta=1/52$.
The axis on the right-hand side displays the time series of 
$\tau_{m,t-\Delta}$.
} \label{maturity 5}
%\end{figure}

%\begin{figure}
\vspace{-1cm}
\begin{center}
\includegraphics[width=\linewidth]{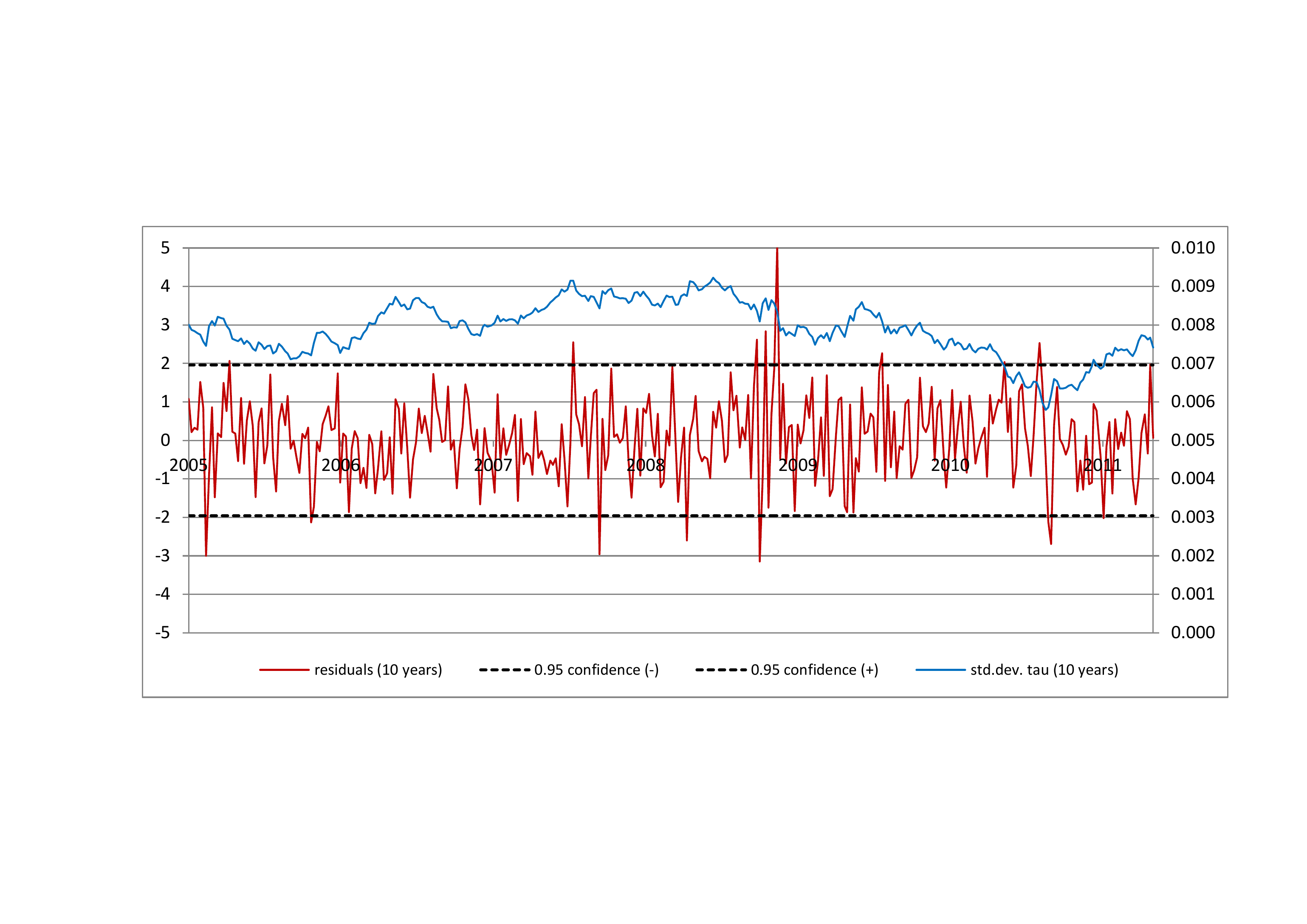}
\end{center}
\vspace{-2cm} 
\caption{Time series  of residuals $z^\ast_{m,t}$
for time to maturity $m=10$ and
$t \in \{01/2005,\ldots, 05/2011\}$
on a weekly grid $\Delta=1/52$.
The axis on the right-hand side displays the time series of 
$\tau_{m,t-\Delta}$.
} \label{maturity 10}
\end{figure}

In order to detect the market price of risk term,
we look at residuals for
individual times to maturity $m\in {\cal M}$, i.e.~we replace the indicators
$\mathbf{1}_{{\cal M}_3}$ in \eqref{life portfolio} by indicators 
$\mathbf{1}_{\{m\}}$. We denote the resulting residuals
by $z_{m,t}^\ast$ and the corresponding volatilities by 
$\tau_{m,t-\Delta}$. In Figures  \ref{maturity 1}, \ref{maturity 5} 
and \ref{maturity 10} we show the results for $m=1,5,10$. 
The picture is similar to Figure \ref{residuals 1}, i.e.~we observe
clustering but not a well-defined drift. This implies that
we suggest to set the market price of risk 
$\boldsymbol{\lambda}_{t}=0$ for the prediction of future 
yield curves (we come back to this in Section \ref{Section Vasicek}).

\subsection{Comparison to the Vasi\v{c}ek model}
\label{Section Vasicek}
We compare our findings to the results in the Vasi\v{c}ek model \cite{Vasicek}.
The Vasi\v{c}ek model is
the simplest short rate model that 
provides an affine term structure for interest rates (see also
Filipovi\'c \cite{Damir}), and hence a closed-form solution
for ZCB prices. The price of the ZCB in the Vasi\v{c}ek model
takes the following form
\begin{equation*}
P(t,t+m)= \exp \left\{ A(m)-r_t~ B(m) \right\},
\end{equation*}
where the short rate process $(r_t)_t$ evolves as an Ornstein-Uhlenbeck
process under $\p^\ast$,
and $A(m)$ and $B(m)$ are constants only depending on the
time to maturity $m$ and the model parameters 
$\kappa^\ast$, $\theta^\ast$ and $g$
(see for instance (3.8) in Brigo-Mercurio \cite{BM}).
The short rate $r_t$ is then 
under $\p^\ast_{t-\Delta}$  normally distributed
with conditional mean and conditional variance given by
\begin{eqnarray*}
\E^\ast_{t-\Delta}[r_t]
&=&r_{t-\Delta}~e^{-\Delta \kappa^\ast}
+\theta^\ast\left(1-e^{-\Delta \kappa^\ast}\right),\\
{\rm Var}^\ast_{t-\Delta}(r_t)&=&
\frac{g^2}{2\kappa^\ast}\left[1-e^{-2\kappa^\ast\Delta}\right].
\end{eqnarray*}
Thus, the approximation $\widetilde{\pi}_t$ has under $\p^\ast_{t-\Delta}$
a normal distribution with conditional mean
\begin{equation*}
\E^\ast_{t-\Delta}[\widetilde{\pi}_t]
= \sum_{m\in {\cal M}_3} \left(1 + A(m) - \E^\ast_{t-\Delta}[r_t] ~B(m)\right),
\end{equation*}
and conditional variance
\begin{equation*}
{\rm Var}^\ast_{t-\Delta}(\widetilde{\pi}_t)
= {\rm Var}^\ast_{t-\Delta}(r_t)
\left(\sum_{m\in {\cal M}_3}  B(m)\right)^2.
\end{equation*}
As in the previous section we 
assume $\p^\ast=\p$, i.e.~we set the market price of risk
$\boldsymbol{\lambda}_{t}=0$: 
(i) this allows to estimate the model parameters
$\kappa^\ast$, $\theta^\ast$ 
and $g$, for instance, using maximum likelihood
methods (see (3.14)-(3.16) in Brigo-Mercurio \cite{BM}); (ii) makes
the model comparable to the calibration of our model. We will comment on this ``comparability'' below.

Thus we estimate these
parameters and obtain parameter estimates $\widehat{\kappa}^\ast$, 
$\widehat{\theta}^\ast$ and $\widehat{g}$ from which
we get the estimated
functions
$\widehat{A}(\cdot)$ and $ \widehat{B}(\cdot)$.
This then allows to estimate  the conditional mean and variance of 
$\widetilde{\pi}_t$, given ${\cal F}_{t-\Delta}$.
From these we calculate the observable residuals
\begin{equation*}
v^\ast_t = \frac{\widetilde{\pi}_t-\widehat{\E}^\ast_{t-\Delta}[\widetilde{\pi}_t]}{
\widehat{{\rm Var}}^\ast_{t-\Delta}(\widetilde{\pi}_t)^{1/2}}.
\end{equation*}
In Figure \ref{residuals_3} we plot the time series
$z^\ast_t$ and $v^\ast_t$ for $t \in \{01/2005,\ldots, 05/2011\}$.
The observation is that $v^\ast_t$
is far too small! The explanation for this observation lies  in the assumption
$\p^\ast=\p$, i.e.~$\boldsymbol{\lambda}_{t}=0$. Since the Vasi\v{c}ek
prices are calculated by conditional  expectations of the 
{\it entire} future
development of the short rate $r_t$ until expiry of the ZCB, the
choice of the market price of risk $\boldsymbol{\lambda}_{t}$
has a huge influence on the
resulting ZCB price in the Vasi\v{c}ek model. Thus, the calibration of
$\widehat{A}(\cdot)$ and $\widehat{B}(\cdot)$ is completely
wrong if we set $\boldsymbol{\lambda}_{t}=0$.
Compare
\begin{eqnarray}
\log P(t,t+m)&=& - m~ Y(t,t+m),\label{ZCB 1}\\
\log P(t,t+m)&=& A(m) - r_t ~B(m) \label{ZCB 2}.
\end{eqnarray}
Conditionally, given ${\cal F}_{t-\Delta}$, we model the development
from $Y(t-\Delta, t+m)$ to $Y(t, t+m)$ for the study of 
\eqref{ZCB 1}. That is, we model a change of the yield curve 
$\mathbf{Y}_{t-\Delta}$ at time $t-\Delta$ to
$\mathbf{Y}_{t}$ at time $t$. Since the yield curve $\mathbf{Y}_{t-\Delta}$
already corresponds to market prices it already contains the actual 
market risk aversion, and thus the market price of risk
$\boldsymbol{\lambda}_t$ in \eqref{market-price of risk}
only influences one single period in our consideration.

The (pricing) functions $A(\cdot)$ and $B(\cdot)$ in \eqref{ZCB 2}, however,
are calculated completely within the Vasi\v{c}ek model by a forward projection
of $r_t$ until maturity date $t+m$. If this forward projection is done
under the wrong measure $\p$, then these pricing components 
completely miss the
market risk aversion and hence are not appropriate. Thus, in general,
we should have $A(m)=A(m,\boldsymbol{\lambda}_{t})$ 
and $B(m)=B(m,\boldsymbol{\lambda}_{t})$ which
requires a detailed knowledge of the market price of risk $\boldsymbol{\lambda}_{t}$
and, thus, the Vasi\v{c}ek model reacts much more sensitively to non-appropriately
calibrated equivalent martingale measures $\p^\ast$. Note that this
is true for all models where ZCB prices are entirely determined
by the short rate process $(r_t)_t$.

\begin{conclusions}~\end{conclusions}
\begin{itemize}
\item
We conclude that the HJM models (similar to Model Assumptions 
\ref{Model Assumptions 1})
are much more robust against inappropriate choices of the market price
of risk compared to short rate models, because in the former we only need
to choose the market price of risk for the one-step ahead for the prediction
of the ZCB prices at the end of the period (i.e.~from $t-\Delta$ to $t$)
whereas for short rate models we need to choose the market price of
risk appropriately for the entire life time of the ZCB (i.e.~from $t-\Delta$
to $t+m$).
\item Our HJM model (Model Assumptions \ref{Model Assumptions 1}) 
always captures the actual yield curve, whereas this is not 
necessarily the case for  short rate models.
\end{itemize}

\subsection{Forward projection of yield curves and arbitrage}
\label{sec.arbitrage}
For the calibration of the model and for yield curve prediction we
have chosen a restricted set ${\cal M}$ of times to maturity.
In most applied cases one has to stay within such a restricted
set because there do not exist observations for all times to
maturity. We propose that we predict future yield
curves within these families ${\cal M}$ and then approximate
the remaining times to maturity using a parametric family
like the Nelson-Siegel \cite{NelsonSiegel}
or the Svensson \cite{Svensson1, Svensson2}
family, see also Filipovi\'c \cite{Damir}.

~

\begin{figure}[ht]
\vspace{-1cm}
\begin{center}
\includegraphics[width=\linewidth]{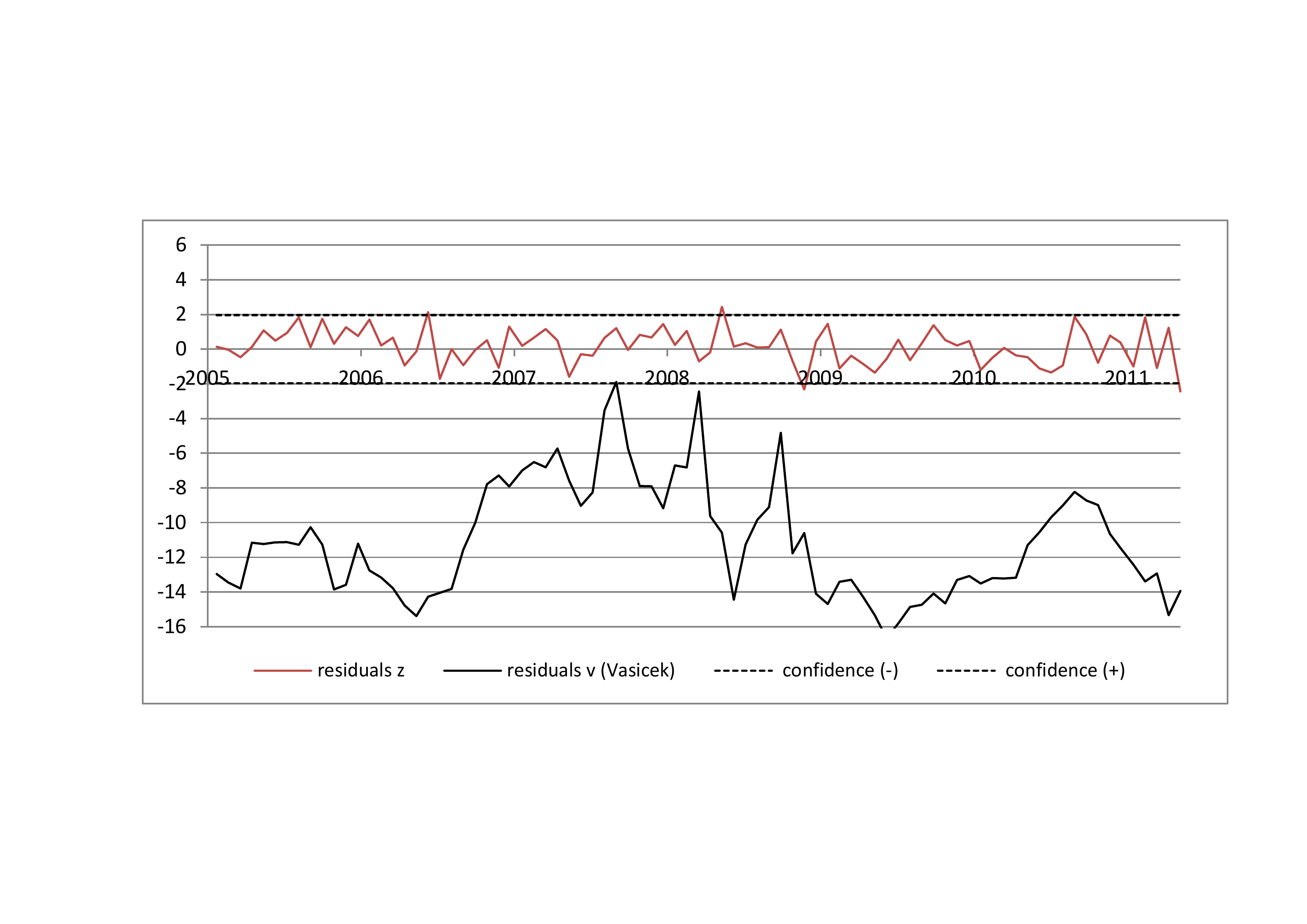}
\end{center}
\vspace{-2cm} 
\caption{Time series  of residuals $z^\ast_t$
and $v^\ast_t$
for $t \in \{01/2005,\ldots, 05/2011\}$
on a monthly grid $\Delta=1/12$ under
the assumption $\p^\ast=\p$.
} \label{residuals_3}
%\end{figure}

%\begin{figure}[ht]
\vspace{-1cm}
\begin{center}
\includegraphics[width=\linewidth]{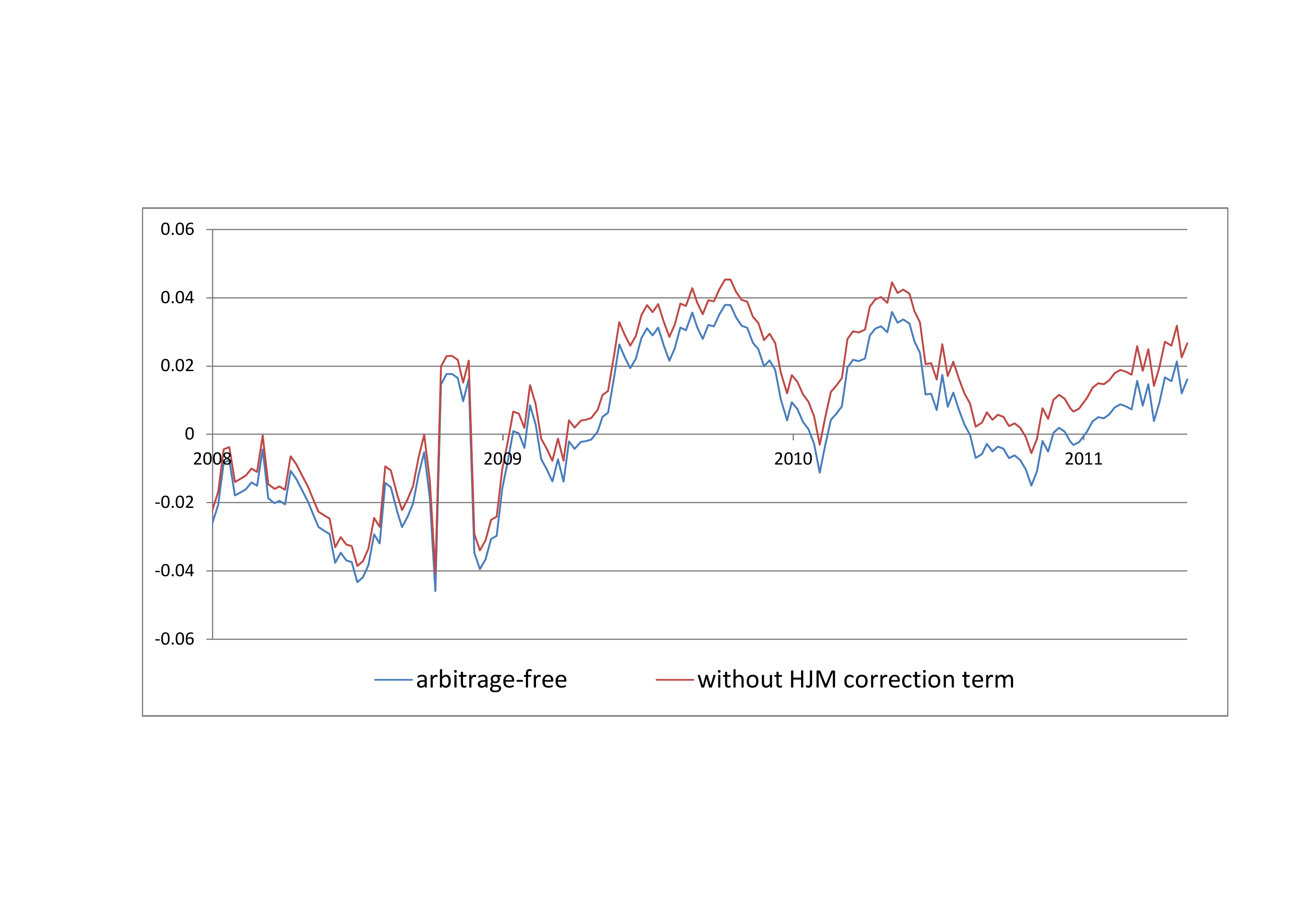}
\end{center}
\vspace{-2cm} 
\caption{Back testing the difference of aggregated realized gains
of portfolio
$\widetilde{\pi}_t$
for $w_t=\tau_{t-\Delta}^{(2)}/\tau_{t-\Delta}^{(1)}$ and the their model prognosis
with and without the no-arbitrage HJM correction term.
} \label{arbitrage}
\end{figure}

Finally, we demonstrate the absence of arbitrage condition given
in Lemma \ref{HJM yield condition}. At the end of
Section \ref{section introduction} we have emphasized the importance
of the no-arbitrage property of the prediction model. Let us choose an
asset portfolio $w_t P(t,t+m_1)-P(t,t+m_2)$ for two different times
to maturity $m_1$ and $m_2$. We approximate this portfolio
by a Taylor expansion up to order $2$ and set
{\small
\begin{equation*}
\widetilde{\pi}_t=
w_t \left(1-m_1Y(t,t+m_1)+\frac{{(m_1Y(t,t+m_1))}^2}{2} \right)-\left(1-m_2Y(t,t+m_2)+\frac{{(m_2Y(t,t+m_2))}^2}{2}\right).
\end{equation*}}
Under our model assumptions, the returns of both terms 
$m_iY(t,t+m_i)$
in portfolio
$\widetilde{\pi}_t$ have, conditionally given ${\cal F}_{t-\Delta}$,
a Gaussian distribution term with standard deviations given by
\begin{equation*}
\tau_{t-\Delta}^{(i)}=
\sqrt{
\mathbf{1}_{\{m_i\}}'~
S_{(K)}
(\mathbf{Y}_{t,-})~\mathbf{1}_{\{m_i\}}}
\qquad \text{ for } i=1,2.
\end{equation*}
If we choose $w_t=\tau_{t-\Delta}^{(2)}/\tau_{t-\Delta}^{(1)}$
then the returns of the Gaussian parts of both terms in portfolio $\widetilde{\pi}_t$ have the same
variance and, thus, under the Gaussian assumption have the same
marginal distributions. Since the conditional expectation of the second order term in the Taylor expansion cancels the no-arbitrage drift term (up to a small short rate correction) we see that the returns of the portfolio $ \widetilde{\pi}_t $ should provide zero returns conditionally.
In Figure \ref{arbitrage} we give an example for times to 
maturity $m_1=10$ and $m_2=20$. The correlation between the
prices of these ZCBs is high, about 85\%, i.e.~their prices tend
to move simultaneously. The resulting weights $w_t$ are in the
range between 1.4 and 1.9. In Figure \ref{arbitrage} we plot the aggregated
realized gains of the portfolio $ \widetilde{\pi} $ minus their prognosis including
and excluding the HJM correction term. Recall that the predicted
gains should be zero conditionally on the current information. We observe that the model without the HJM term
clearly drifts away from zero, which opens the possibility of arbitrage. Therefore, we insist on a prediction model that is free of arbitrage.

\appendix

\section{Proofs}
{\Beweis
{\bf Proof of Theorem \ref{theorem unbiased}.}
In the first step we apply the tower property for conditional
expectation which decouples the problem into several steps. We
have $\E^\ast_0 \left[S_{(K)}(\mathbf{y})\right]
=\E^\ast_0 \left[\E^\ast_{\Delta(K-1)}\left[S_{(K)}(\mathbf{y})\right]\right]$.
Thus, we need to calculate the inner conditional expectation 
$\E^\ast_{\Delta(K-1)}\left[\cdot\right]$ of the 
$d\times d$ matrix $S_{(K)}(\mathbf{y})$. We define
the auxiliary matrix
\begin{equation*}
\widetilde{C}_{(K)} = \left(\left[
\varsigma(\mathbf{Y}_{\Delta k,-})^{-1}~
\boldsymbol{\Upsilon}_{\Delta k}\right]_j
\right)_{j=1,\ldots, d; ~k=1, \ldots, K}\in \R^{d\times K}.
\end{equation*}
This implies that we can rewrite $
C_{(K)}= K^{-1/2}~\widetilde{C}_{(K)}$.
Moreover, we rewrite the matrix $\widetilde{C}_{(K)}$
as follows
\begin{equation*}
\widetilde{C}_{(K)}= \left[\widetilde{C}_{(K-1)},
\varsigma(\mathbf{Y}_{\Delta K,-})^{-1}~
\boldsymbol{\Upsilon}_{\Delta K} \right],
\end{equation*}
with $\widetilde{C}_{(K-1)}\in \R^{d\times(K-1)}$ is 
${\cal F}_{\Delta(K-1)}$-measurable.
This implies the following decomposition
\begin{eqnarray*}
S_{(K)}(\mathbf{y})
&=&\frac{1}{K}~
\varsigma(\mathbf{y})~\widetilde{C}_{(K)}
~\widetilde{C}_{(K)}'~\varsigma(\mathbf{y})'\\
&=&\frac{1}{K}~
\varsigma(\mathbf{y})\left[\widetilde{C}_{(K-1)},
\varsigma(\mathbf{Y}_{\Delta K,-})^{-1}~
\boldsymbol{\Upsilon}_{\Delta K} \right]
\left[\widetilde{C}_{(K-1)},
\varsigma(\mathbf{Y}_{\Delta K,-})^{-1}~
\boldsymbol{\Upsilon}_{\Delta K} \right]'\varsigma(\mathbf{y})'\\
&=&\frac{1}{K}~
\varsigma(\mathbf{y})
\left(\widetilde{C}_{(K-1)}~\widetilde{C}'_{(K-1)}
+
\left(\varsigma(\mathbf{Y}_{\Delta K,-})^{-1}~
\boldsymbol{\Upsilon}_{\Delta K} \right)
\left(\varsigma(\mathbf{Y}_{\Delta K,-})^{-1}~
\boldsymbol{\Upsilon}_{\Delta K} \right)'\right)\varsigma(\mathbf{y})'
\\
&=&\frac{K-1}{K}~S_{(K-1)}(\mathbf{y})
+\frac{1}{K}~
\varsigma(\mathbf{y})~
\varsigma(\mathbf{Y}_{\Delta K,-})^{-1}~
\boldsymbol{\Upsilon}_{\Delta K}~ 
\boldsymbol{\Upsilon}_{\Delta K}'
\left(\varsigma(\mathbf{Y}_{\Delta K,-})^{-1}\right)'~
\varsigma(\mathbf{y})'.
\end{eqnarray*}
This implies for the conditional expectation of $S_{(K)}(\mathbf{y})$
\begin{equation*}
\E^\ast_{\Delta(K-1)}\left[S_{(K)}(\mathbf{y})\right]
=\frac{K-1}{K}~S_{(K-1)}(\mathbf{y})
%\\&&
+\frac{1}{K}~
\varsigma(\mathbf{y})~
\varsigma(\mathbf{Y}_{\Delta K,-})^{-1}~
\E^\ast_{\Delta(K-1)}\left[
\boldsymbol{\Upsilon}_{\Delta K}~ 
\boldsymbol{\Upsilon}_{\Delta K}'\right]
\left(\varsigma(\mathbf{Y}_{\Delta K,-})^{-1}\right)'~
\varsigma(\mathbf{y})'.
\end{equation*}
We calculate the conditional expectation in the last term,
we start with the conditional covariance. From
Lemma \ref{Lemma 3.2} we obtain
\begin{eqnarray*}
%&&\hspace{-1cm}
\frac{1}{K}~
\varsigma(\mathbf{y})~
\varsigma(\mathbf{Y}_{\Delta K,-})^{-1}~
{\rm Cov}^\ast_{\Delta(K-1)}\left(
\boldsymbol{\Upsilon}_{\Delta K}\right)
\left(\varsigma(\mathbf{Y}_{\Delta K,-})^{-1}\right)'~
\varsigma(\mathbf{y})'\\=~
\frac{\Delta}{K}~
\varsigma(\mathbf{y})~
\varsigma(\mathbf{Y}_{\Delta K,-})^{-1}~
\Sigma_{\Lambda} 
({\mathbf{Y}_{\Delta K, -}})
\left(\varsigma(\mathbf{Y}_{\Delta K,-})^{-1}\right)'~
\varsigma(\mathbf{y})'
&=&
\frac{\Delta}{K}~
\Sigma_{\Lambda} (\mathbf{y}).
\end{eqnarray*}
%Thus, from 
%\begin{equation*}
%\E^\ast_{\Delta(K-1)}\left[
%\boldsymbol{\Upsilon}_{\Delta K}~ 
%\boldsymbol{\Upsilon}_{\Delta K}'\right]
%=
%{\rm Cov}^\ast_{\Delta(K-1)}\left(
%\boldsymbol{\Upsilon}_{\Delta K}\right)+
%\E^\ast_{\Delta(K-1)}\left[
%\boldsymbol{\Upsilon}_{\Delta K}\right]~ 
%\E^\ast_{\Delta(K-1)}\left[
%\boldsymbol{\Upsilon}_{\Delta K}\right]',
%\end{equation*}
%
%there remains the calculation of the last term.
This implies
\begin{eqnarray*}
\E^\ast_{0}\left[S_{(K)}(\mathbf{y})\right]
&=&\frac{K-1}{K}~\E^\ast_{0}\left[S_{(K-1)}(\mathbf{y})\right]+
\frac{\Delta}{K}~
\Sigma_{\Lambda} (\mathbf{y})
\\&&
+\frac{1}{K}~
\varsigma(\mathbf{y})~
\E^\ast_{0}\left[
\varsigma(\mathbf{Y}_{\Delta K,-})^{-1}~
\E^\ast_{\Delta(K-1)}\left[
\boldsymbol{\Upsilon}_{\Delta K}\right]~ 
\E^\ast_{\Delta(K-1)}\left[\boldsymbol{\Upsilon}_{\Delta K}\right]'
\left(\varsigma(\mathbf{Y}_{\Delta K,-})^{-1}\right)'\right]~
\varsigma(\mathbf{y})'\\
&=&\frac{K-1}{K}~\E^\ast_{0}\left[S_{(K-1)}(\mathbf{y})\right]+
\frac{\Delta}{K}~
\Sigma_{\Lambda} (\mathbf{y})
+\frac{\Delta^2}{K}~
\varsigma(\mathbf{y})~
\E^\ast_{0}\left[
f_\Lambda(\mathbf{Y}(\Delta(K-1),\Delta K),\mathbf{Y}_{\Delta K, -})\right]~
\varsigma(\mathbf{y})'.
\end{eqnarray*}
Iterating this provides the result.
\EndProof}

\newpage

\begin{landscape}
\begin{center}
\begin{table}[ht]
{\scriptsize
\begin{tabular}{c|rrrr|rrrrrrrrrrrrr}
\hline
           &     1 week &    2 weeks &    1M  &   3M &     1Y &    2Y &    
3Y &    4Y &    5Y &    6Y &    7Y &    8Y &    9Y &   10Y &   15Y 
&   20Y &   30Y \\
\hline
    1 week &     0.0000 &     0.0001 &     0.0001 &     0.0002 &     0.0004 &     0.0005 &     0.0005 &     0.0006 &     0.0006 &     0.0007 &     0.0007 &     0.0007 &     0.0007 &     0.0007 &     0.0009 &     0.0011 &     0.0017 \\

   2 weeks &     0.0001 &     0.0001 &     0.0002 &     0.0003 &     0.0005 &     0.0006 &     0.0006 &     0.0007 &     0.0008 &     0.0008 &     0.0008 &     0.0009 &     0.0009 &     0.0009 &     0.0011 &     0.0014 &     0.0023 \\

   1 month &     0.0001 &     0.0002 &     0.0004 &     0.0005 &     0.0005 &     0.0007 &     0.0008 &     0.0009 &     0.0009 &     0.0009 &     0.0009 &     0.0009 &     0.0009 &     0.0009 &     0.0010 &     0.0016 &     0.0028 \\

  3 months &     0.0002 &     0.0003 &     0.0005 &     0.0028 &     0.0019 &     0.0024 &     0.0021 &     0.0018 &     0.0017 &     0.0016 &     0.0017 &     0.0017 &     0.0018 &     0.0019 &     0.0025 &     0.0042 &     0.0064 \\
\hline
    1 year &     0.0004 &     0.0005 &     0.0005 &     0.0019 &     0.0142 &     0.0164 &     0.0158 &     0.0161 &     0.0168 &     0.0173 &     0.0176 &     0.0176 &     0.0175 &     0.0173 &     0.0173 &     0.0208 &     0.0371 \\

   2 years &     0.0005 &     0.0006 &     0.0007 &     0.0024 &     0.0164 &     0.0309 &     0.0361 &     0.0373 &     0.0376 &     0.0380 &     0.0388 &     0.0397 &     0.0410 &     0.0423 &     0.0500 &     0.0590 &     0.0658 \\

   3 years &     0.0005 &     0.0006 &     0.0008 &     0.0021 &     0.0158 &     0.0361 &     0.0481 &     0.0535 &     0.0562 &     0.0581 &     0.0600 &     0.0619 &     0.0641 &     0.0663 &     0.0779 &     0.0907 &     0.1019 \\

   4 years &     0.0006 &     0.0007 &     0.0009 &     0.0018 &     0.0161 &     0.0373 &     0.0535 &     0.0632 &     0.0693 &     0.0735 &     0.0771 &     0.0801 &     0.0830 &     0.0857 &     0.0990 &     0.1126 &     0.1327 \\

   5 years &     0.0006 &     0.0008 &     0.0009 &     0.0017 &     0.0168 &     0.0376 &     0.0562 &     0.0693 &     0.0787 &     0.0851 &     0.0904 &     0.0947 &     0.0986 &     0.1021 &     0.1173 &     0.1310 &     0.1581 \\

   6 years &     0.0007 &     0.0008 &     0.0009 &     0.0016 &     0.0173 &     0.0380 &     0.0581 &     0.0735 &     0.0851 &     0.0936 &     0.1007 &     0.1065 &     0.1116 &     0.1161 &     0.1342 &     0.1485 &     0.1786 \\

   7 years &     0.0007 &     0.0008 &     0.0009 &     0.0017 &     0.0176 &     0.0388 &     0.0600 &     0.0771 &     0.0904 &     0.1007 &     0.1095 &     0.1169 &     0.1234 &     0.1292 &     0.1520 &     0.1674 &     0.1969 \\

   8 years &     0.0007 &     0.0009 &     0.0009 &     0.0017 &     0.0176 &     0.0397 &     0.0619 &     0.0801 &     0.0947 &     0.1065 &     0.1169 &     0.1259 &     0.1340 &     0.1413 &     0.1700 &     0.1871 &     0.2135 \\

   9 years &     0.0007 &     0.0009 &     0.0009 &     0.0018 &     0.0175 &     0.0410 &     0.0641 &     0.0830 &     0.0986 &     0.1116 &     0.1234 &     0.1340 &     0.1438 &     0.1528 &     0.1883 &     0.2078 &     0.2297 \\

  10 years &     0.0007 &     0.0009 &     0.0009 &     0.0019 &     0.0173 &     0.0423 &     0.0663 &     0.0857 &     0.1021 &     0.1161 &     0.1292 &     0.1413 &     0.1528 &     0.1635 &     0.2064 &     0.2289 &     0.2462 \\

  15 years &     0.0009 &     0.0011 &     0.0010 &     0.0025 &     0.0173 &     0.0500 &     0.0779 &     0.0990 &     0.1173 &     0.1342 &     0.1520 &     0.1700 &     0.1883 &     0.2064 &     0.2869 &     0.3320 &     0.3498 \\

  20 years &     0.0011 &     0.0014 &     0.0016 &     0.0042 &     0.0208 &     0.0590 &     0.0907 &     0.1126 &     0.1310 &     0.1485 &     0.1674 &     0.1871 &     0.2078 &     0.2289 &     0.3320 &     0.4247 &     0.5215 \\

  30 years &     0.0017 &     0.0023 &     0.0028 &     0.0064 &     0.0371 &     0.0658 &     0.1019 &     0.1327 &     0.1581 &     0.1786 &     0.1969 &     0.2135 &     0.2297 &     0.2462 &     0.3498 &     0.5215 &     0.9860 \\
\hline
\end{tabular}  }
\caption{Estimated matrix $\widehat{\Sigma}_\Lambda(\mathbf{1})
=(\widehat{s}_{ij}(K))_{i,j=1,\ldots, d}$ based on all observations 
in $ \{01/2000,\ldots, 05/2011\}$
on a weekly
grid $\Delta=1/52$.} \label{Table 1}
\end{table}

\begin{table}[ht]
{\scriptsize
\begin{center}
\begin{tabular}{c|r|rrrrrrrrrrrrr}
\hline
           &   3M  &     1Y &    2Y &    3Y &    4Y &    5Y &    6Y 
&    7Y &    8Y &    9Y &   10Y &   15Y &   20Y &   30Y \\
\hline

3 months&      31\% &       70\% &       72\% &       78\% &       82\% &       84\% &       85\% &       85\% &       85\% &       85\% &       84\% &       83\% &       76\% &       76\% \\
\hline
1 year&      70\% &        2\% &        7\% &       16\% &       19\% &       21\% &       23\% &       24\% &       27\% &       29\% &       30\% &       38\% &       36\% &       24\% \\

2 years&      72\% &        7\% &       -6\% &       -1\% &        6\% &       11\% &       13\% &       14\% &       15\% &       16\% &       15\% &       15\% &       13\% &       23\% \\

3 years &      78\% &       16\% &       -1\% &        0\% &        2\% &        4\% &        6\% &        7\% &        7\% &        8\% &        7\% &        8\% &        7\% &       15\% \\

4 years&      82\% &       19\% &        6\% &        2\% &        1\% &        1\% &        1\% &        2\% &        2\% &        3\% &        3\% &        6\% &        7\% &       11\% \\

5 years&      84\% &       21\% &       11\% &        4\% &        1\% &       -1\% &       -1\% &       -1\% &       -1\% &        0\% &        1\% &        5\% &        8\% &       11\% \\

6 years&      85\% &       23\% &       13\% &        6\% &        1\% &       -1\% &       -2\% &       -2\% &       -2\% &       -1\% &        0\% &        5\% &        9\% &       12\% \\

7 years&      85\% &       24\% &       14\% &        7\% &        2\% &       -1\% &       -2\% &       -3\% &       -2\% &       -1\% &       -1\% &        4\% &        9\% &       14\% \\

8 years&      85\% &       27\% &       15\% &        7\% &        2\% &       -1\% &       -2\% &       -2\% &       -2\% &       -1\% &       -1\% &        3\% &        9\% &       16\% \\

9 years&      85\% &       29\% &       16\% &        8\% &        3\% &        0\% &       -1\% &       -1\% &       -1\% &       -1\% &       -1\% &        2\% &        8\% &       18\% \\

10 years&      84\% &       30\% &       15\% &        7\% &        3\% &        1\% &        0\% &       -1\% &       -1\% &       -1\% &       -1\% &        1\% &        7\% &       20\% \\

15 years&      83\% &       38\% &       15\% &        8\% &        6\% &        5\% &        5\% &        4\% &        3\% &        2\% &        1\% &        0\% &        5\% &       20\% \\

20 years&      76\% &       36\% &       13\% &        7\% &        7\% &        8\% &        9\% &        9\% &        9\% &        8\% &        7\% &        5\% &        3\% &        7\% \\

30 years&      76\% &       24\% &       23\% &       15\% &       11\% &       11\% &       12\% &       14\% &       16\% &       18\% &       20\% &       20\% &        7\% &      -24\% \\

\hline
\end{tabular}  
\end{center}
}
\caption{Estimated matrices $\widehat{\Sigma}_\Lambda(\mathbf{1})
=(\widehat{s}_{ij}(K))_{i,j=1,\ldots, d}$ based on all observations 
in $ \{01/2000,\ldots, 05/2011\}$. The table shows the differences
between the estimates on a weekly grid $\Delta=1/52$ versus the
estimates on a quarterly grid $\Delta =1/4$ (relative to the 
estimated values on the quarterly grid).} \label{Table 2}
\end{table}

\end{center}
\end{landscape}


{\small 
\begin{thebibliography}{99}

\bibitem{BM}
Brigo, D., Mercurio, F. (2006).
Interest Rate Models - Theory and Practice.
Springer.


\bibitem{DS}
Delbaen, F., Schachermayer, W. (1994).
A general version of the fundamental theorem of 
asset pricing. Mathematische Annalen 300, 463-520.


\bibitem{Damir}
Filipovi\'c, D. (2009).
Term-Structure Models. Springer.

\bibitem{HJM}
Heath, D., Jarrow, R., Morton, A. (1992).
Bond pricing and the term structure of interest rates:
a new methodology. Econometrica 60, 77-105.

\bibitem{Jordan}
Jordan, T.J. (2009).
SARON - an innovation for the financial markets.
Launch event for Swiss Reference Rates, Zurich, August 25, 2009.

\bibitem{NelsonSiegel}
Nelson, C.R., Siegel, A.F. (1987).
Parsimonious modeling of yield curves.
J.~Business 60/4, 473-489.


\bibitem{Ortega}
Ortega, J.P., Pullirsch, R., Teichmann, J., Wergieluk, J. (2009).
A dynamic approach for scenario generation in risk management.
Preprint on arXiv.

\bibitem{Svensson1}
Svensson, L.E.O. (1994).
Estimating and interpreting forward interest rates:
Sweden 1992-1994.
NBER Working Paper Series Nr.~4871.


\bibitem{Svensson2}
Svensson, L.E.O. (1995).
Estimating forward interest rates with
the extended Nelson \& Siegel method.
Sveriges Riksbank Quarterly Review 3, 13-26.

\bibitem{Vasicek}
Vasi\v{c}ek, O. (1977).
An equilibrium characterization of the term structure.
J.~Financial Economics 5/2, 177-188.


\end{thebibliography}}
\end{document}